\PassOptionsToPackage{unicode}{hyperref}
\documentclass[titlepage]{article}

\usepackage{ifthen}
\newboolean{internal}
\setboolean{internal}{false}
\newboolean{arxiv}
\setboolean{arxiv}{true}
\ifthenelse{\boolean{internal}}{
  \typeout{internal}
  \def\ds{}
  \def\dodouble{}
  \def\attach{true}
  
}{
  \ifthenelse{\boolean{arxiv}}{
    \typeout{arxiv}
    \def\ds{}
    \def\dodouble{}
    \def\attach{false}
    
  }{
    \typeout{external}
    \def\ds{doublespacing}
    \def\dodouble{\doublespacing}
    \def\attach{false}
    
  }
}

\usepackage{amstext}
\usepackage[T1]{fontenc}
\usepackage{textgreek}
\usepackage{allbib}

\usepackage{doiref}
\usepackage{ifoo}
\ifoo{
  \usepackage[nomarkers,nolists]{endfloatfix}
  \usepackage[attach=false]{movieref}
}{
  \usepackage[attach=\attach]{movieref}
}

\usepackage{setspace}
\usepackage[font={\ds,small}]{caption}

\usepackage[round]{natbib}
\usepackage{amsmath}

\usepackage{subfig}

\usepackage{underscore}
\usepackage{multirow}
\usepackage{tabularx}

\usepackage{graphicx}
\usepackage{titlepage}
\usepackage{supplementary}
\usepackage{section}

\title[Electrostatics of voltage sensitivity]{
  Electrostatic determinants of voltage sensitivity in ion channels:\\
  Simulations of sliding-helix mechanisms
}


\author{
  Alexander Peyser\footnote\department
  \and Wolfgang Nonner\footnotemark[1]
}
\def\department{Department of Physiology and Biophysics, University of Miami}
\address{
  Alexander Peyser\\
  P.O. Box 016430\\
  Miami, FL  33101-R430\\
  \\
  Phone: 786-877-7398\\
  Fax: 305-243-5931\\
  Email: peyser.alex@gmail.com
}

\begin{document}
\ifoo{\newcommand\dbl[3]{{{#1}_{#2}}^{#3}}}{\newcommand\dbl[3]{{#1}_{#2}^{#3}}}

\maketitle\dodouble

\begin{abstract}
  Electrical signaling via voltage-gated ion channels depends upon the
  function of the voltage sensor (VS), identified with the S1--S4
  domain of voltage-gated K channels. Here we investigate some
  physical aspects of the sliding-helix model of the VS using
  simulations based on VS charges, linear dielectrics and whole-body
  motion. Model electrostatics in voltage-clamped boundary conditions
  are solved using a boundary element method. The statistical
  mechanical consequences of the electrostatic configurational energy
  are computed to gain insight into the sliding-helix mechanism and to
  predict experimentally measured ensemble properties such as gating
  charge displaced by an applied voltage. Those consequences and
  ensemble properties are investigated for variations of: S4
  configuration (α- and $3_{10}$-helical), intrinsic counter-charges,
  protein polarizability, geometry of the gating canal, screening of
  S4 charges by the baths, and protein charges located at the bath
  interfaces. We find that the sliding helix VS has an inherent
  electrostatic stability and its function as a VS is robust in the
  parameter space explored. Maximal charge displacement is limited by
  geometry, specifically the range of movement where S4 charges and
  counter-charges overlap in the region of weak dielectric. The
  steepness of charge rearrangement in the physiological voltage range
  is sensitive to the landscape of electrostatic energy: energy
  differences of <2~kT have substantial consequences. Such
  variations of energy landscape are produced by all variations of
  model features tested. The amount of free energy per unit voltage
  that a sliding-helix VS can deliver to other parts of the channel
  (conductance voltage sensitivity) is limited by both the maximal
  displaced charge and the steepness of charge redistribution by
  voltage (sensor voltage sensitivity).
\end{abstract}

\section{Introduction}

Electrical excitability of cells is possible because the movement of a
few charges can control the flow of many charges. This principle ---
amplification --- led \citet{hodgkin:1952:quant} to their theory of
the action potential in terms of electrically controlled membrane
conductances. Electrically controlled conductances have been localized
to channel proteins conducting sodium, potassium or calcium ions
across the cell membrane. Voltage-controlled ionic conductances and
underlying intrinsic charge movements of ion channels have been
studied experimentally for many years (reviewed by
\citealp{hille:2001}). Electrophysiology has been complemented in more
recent years by techniques measuring channel topology, channel
structure and the change in channel function (reviewed by
\citealp{gandhi:2002,catterall:2010}). Together, these perspectives
provide a wealth of detailed information on the `voltage sensor' (VS)
common to these channels and exemplified by the S1--S4 transmembrane
domains of \emph{Shaker}-type $\text{K}^+$ channels. Here we use computation
to identify physical properties that are important for voltage
sensing, with a natural focus on electrostatic properties.

Site-directed mutagenesis experiments have shown that only a limited
number of amino acid residues of a voltage-dependent ion channel are
individually important for voltage sensitivity \citep{gandhi:2002}. To
study VS electrostatics, we use models in which amino acids are not
made explicit. Since amino-acid residues with charged side-chains in
S4 as well as S2 and S3 transmembrane segments strongly determine VS
function \citep{papazian:1995}, we explicitly represent those
charges. These charged groups operate in vivo in an environment
composed of materials of diverse electrical polarizabilities. We
distinguish three such materials in the model (membrane lipid, baths,
and protein) and describe electrical polarizabilities by dielectric
coefficients that are uniform within each region.

The electrostatic model constructed from point charges and piecewise
uniform dielectrics is encapsulated by conductors clamped to imposed
potentials from external sources while external charge flow is
monitored. Thus the model is coupled to the macroscopic electrical
setup of a `voltage clamp' experiment. The electrostatics of the
system composed of the VS model and the electrode setup is solved
self-consistently. The configuration space of the model VS is
systematically sampled to construct a partition function based on
electrostatic configurational energy. Using this partition function,
we compute ensemble expectations of observable random variables (e.g.,
of gating charge displaced at an applied voltage). This way our
theoretical results meet two criteria for practical usefulness: they
are consistent solutions of the physics included in the model, and
they directly pertain to macroscopic experimental results.

The purpose of our simulations is to elucidate how physical features
of the VS determine functional characteristics. To this end, we
determine the consequences of varying electrostatic and geometrical
parameters of VS models.

\section{Model and boundary conditions}

We test VS models in a simulated voltage clamp setup
(Fig.~\ref{fig:cell}, further geometrical information is given in
Table~\ref{tab:cell}).  The simulation cell is represented by an axial
cross-section of the radially symmetric three-dimensional domain swept
by rotating that cross-section about its vertical axis. The external
boundaries (in green) are electrode surfaces kept at controlled
electrical potentials. The blue zones represent aqueous baths (with a
dielectric coefficient $\epsilon_{\mathrm{w}}=80$). The pink zone is a
region of small dielectric coefficient ($\epsilon_{\mathrm{m}}=2$)
that represents the lipid membrane. The brown zone represents the
region of the channel protein that we model; this region is assigned a
dielectric coefficient of $\epsilon_{\mathrm{p}}=4$ unless noted
otherwise. These dielectrics are piecewise uniform and therefore have
sharp boundaries (solid black lines). Point charges representing
protein charges of interest are embedded in the region of protein
dielectric. Variation of their placement is part of this study and
will be detailed later. The protein region as seen here represents the
matrix of the S4 helix as a central cylinder, surrounded by the other
parts of the channel that create as well a dielectric environment
different from the dielectric environment of the membrane
lipid. Included in the `protein' region are the invaginations which
allow the baths to extend into the planes defining the lipid phase of
the membrane. The radius of the S4 dielectric matrix is 1~nm
(α-helix) or 0.98~nm ($3_{10}$-helix).

\begin{figure}
  \mbox{
    \begin{minipage}{0.75\columnwidth}
      \centering
      \subfloat[]{
        \includegraphics{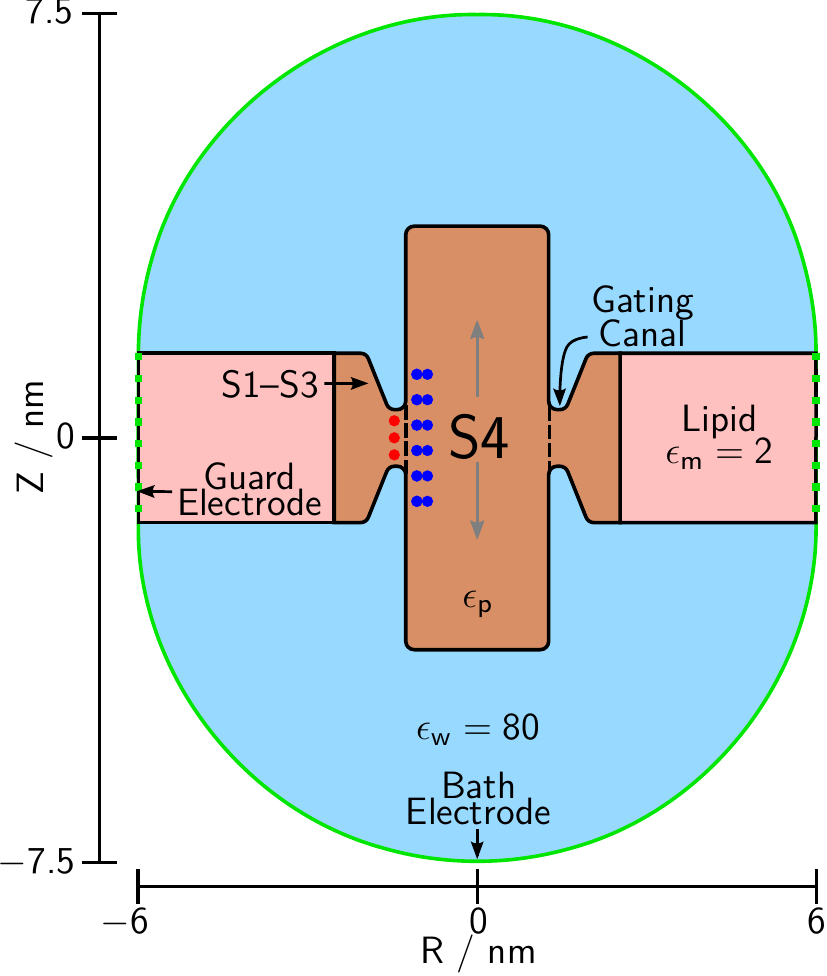}
        \label{fig:cell:both}
      }
    \end{minipage}
    \begin{minipage}{0.25\columnwidth}
      \centering
      \subfloat[]{
        \includegraphics{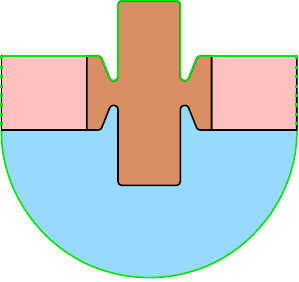}
        \label{fig:cell:bottom}
      }\\
      \subfloat[]{
        \includegraphics{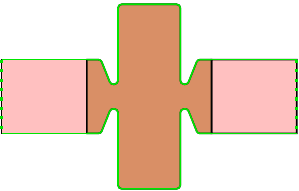}
        \label{fig:cell:none}
      }
    \end{minipage}
  }
  \caption{\emph{Simulation cell.} (A) The 3D setup is produced by
    rotating the cross-section about its vertical axis. \emph{Green
      lines} are electrode surfaces bounding the cell. \emph{black
      lines} are dielectric boundaries separating uniform dielectric
    regions: baths (\emph{blue}), membrane lipids (\emph{pink}), and
    protein (\emph{brown}) with dielectric coefficient
    $\epsilon_{\mathrm{p}}$ (varied between 2 and 16). Charges of
    protein side chains (radial positions represented by color balls
    --- see Fig.~\ref{fig:helix}) are embedded in the protein
    dielectric region in varied geometries. We thus simulate a single
    VS sensor domain (S1--S4) modeled as a central S4 cylinder
    surrounded by a ring of protein material including the S1--S3
    transmembrane domains. The junction between these protein domains
    is narrowed to less than the membrane thickness by circular
    invaginations (`vestibules') leading up to the `gating canal'
    through which the S4 helix glides through the rest of the protein
    (dashed line). Precise lengths defining this geometry are
    specified in Table~\ref{tab:cell}. (B, C) Modified simulation
    cells used for simulating high ionic strength for the external
    bath (B) or for both baths (C) by placing the respective
    electrode(s) directly on the membrane and protein.}
\label{fig:cell}
\end{figure}

Point charges representing protein charges are arranged at a minimal
distance of 0.2~nm from the protein/water boundary. The charged
guanido group of each arginine residue of the S4 segment is
represented as three point charges of 1/3~$\text{e}_0$ on a circle of radius
0.122~nm (blue spheres in Fig.~\ref{fig:helix}). The centers of the S4
arginine charges are arranged on a helix defined by arginine side
chains on an α- or $3_{10}$-helix backbone, where every third
amino acid is an arginine. For the α-helix, charged residues
are separated by 0.45~nm in the transmembrane direction and 60\textdegree
leftward around the helix; for the $3_{10}$-helix, charged residues
are separated by 0.6~nm and 0\textdegree.

Dimensions of the simulation surfaces are varied within those
constraints as defined in Table~\ref{tab:cell}.
Part~\subref*{tab:cell:outline} maps the topology of the protein and
membrane surface of Fig.~\ref{fig:cell} onto metrics for simulations,
defined in part~\subref*{tab:cell:params} of the table. Since the
surfaces in the system are radially symmetrical and smooth, the system
can be defined by a set of inflection points with their curvature on
the left half of the system to be simulated. When the gating canal is
symmetrical, this reduces further to the upper half of the left
side. (As illustrated with
Figs.~\ref{fig:cell}\subref*{fig:cell:both}--\subref*{fig:cell:none}
and discussed later in ``Results'' under ``Screening of VS charges'',
these metrics apply for ``dilute'' systems with electrodes far from
the membrane surface, ``infinitely concentrated'' systems with
electrodes on the membranes, and asymmetrical systems with a dilute
side and a concentrated side.)

\begin{table}
  \centering
  \subfloat[\vskip-2\baselineskip]{
    \label{tab:cell:outline}
    \begin{minipage}{\columnwidth}
      \centering
      \includegraphics[width=0.6\columnwidth]{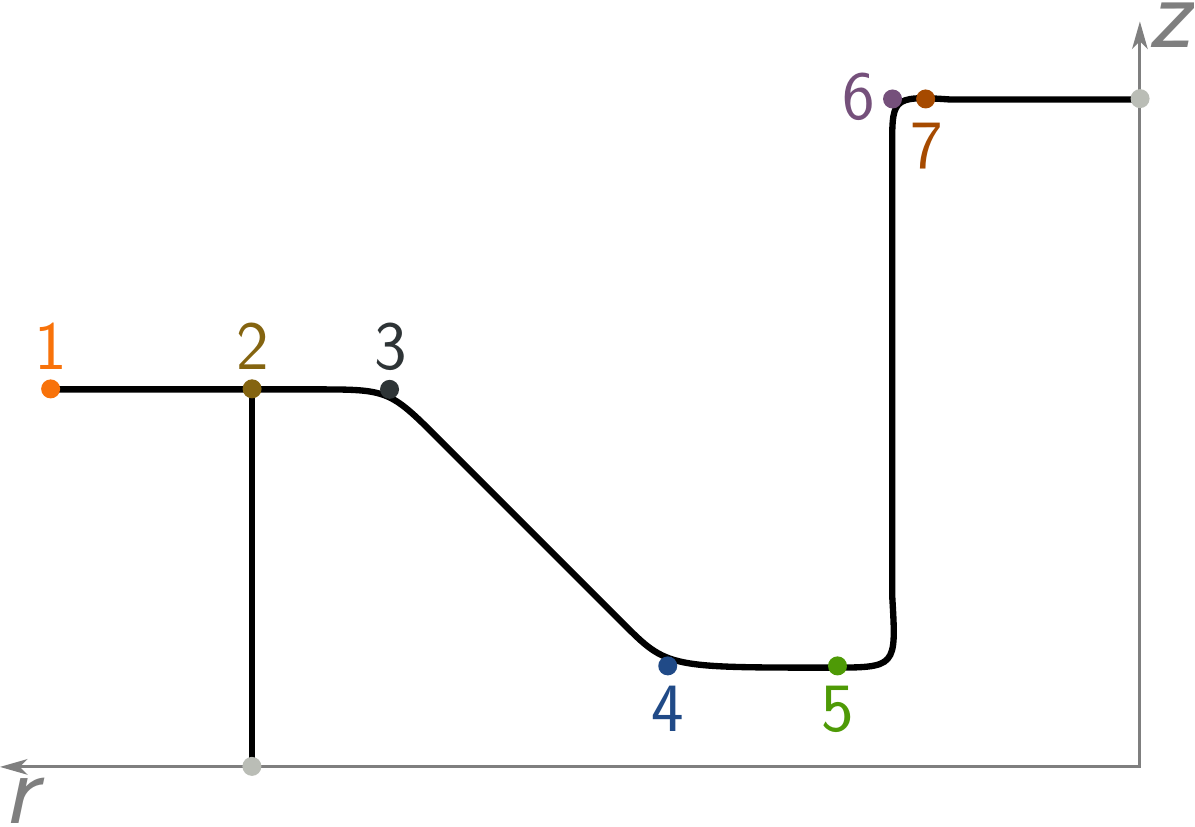}
    \end{minipage}
  }\\
  \subfloat[]{
    \small
    \label{tab:cell:params}
    \def\mr#1#2{\multirow{#1}{*}{\parbox{0.35in}{\baselineskip=0pt\centering #2}}}
    \begin{tabular}{cc|*{7}{r@.l}}
        \multicolumn{2}{c|}{\textbf{Model}}
      & \multicolumn{2}{c}{\textbf{1}}
      & \multicolumn{2}{c}{\textbf{2}}
      & \multicolumn{2}{c}{\textbf{3}}
      & \multicolumn{2}{c}{\textbf{4}}
      & \multicolumn{2}{c}{\textbf{5}}
      & \multicolumn{2}{c}{\textbf{6}}
      & \multicolumn{2}{c}{\textbf{7}} \\\hline
      & \\[-1ex]
      \mr{2}{α\\(1)}
      & $r$
      & 6&0
      & 2&532
      & 1&966
      & 1&566
      & 1&466
      & 1&266
      & 1&0
      \\
      & $z$
      & 1&5
      & 1&5
      & 1&5
      & 0&5015
      & 0&5015
      & 3&7515
      & 3&7515
      \\[1ex]
      \mr{2}{$3_{10}$}
      & $r$
      & 6&0
      & 2&492
      & 1&946
      & 1&546
      & 1&446
      & 1&246
      & 9&8
      \\
      & $z$
      & 1&5
      & 1&5
      & 1&5
      & 0&602
      & 0&602
      & 4&602
      & 4&602
      \\[1ex]
      \mr{3}{α\\(2)}
      & $r$
      & 6&0
      & 2&532
      & 1&966
      & 1&566
      & 1&466
      & 1&266
      & 1&0
      \\
      & $+z$
      & 1&5
      & 1&5
      & 1&5
      & 0&10225
      & 0&10225
      & 3&7515
      & 3&7515
      \\
      & $-z$
      & 1&5
      & 1&5
      & 1&5
      & -0&90075
      & -0&90075
      & 3&7515
      & 3&7515
      \\[1ex]
      \mr{2}{α\\(3)}
      & $r$
      & 6&0
      & 2&532
      & 1&966
      & 1&566
      & 1&466
      & 1&266
      & 1&0
      \\
      & $z$
      & 1&5
      & 1&5
      & 1&5
      & 0&90075
      & 0&90075
      & 4&15075
      & 4&15075
      \\[1ex]
      \mr{3}{α\\(4)}
      & $r$
      & 6&0
      & 2&532
      & 1&966
      & 1&566
      & 1&466
      & 1&266
      & 1&0
      \\
      & $+z$
      & 1&5
      & 1&5
      & 1&5
      & 0&5015
      & 0&5015
      & 4&15075
      & 4&15075
      \\
      & $-z$
      & 1&5
      & 1&5
      & 1&5
      & -1&3
      & -1&3
      & 4&15075
      & 4&15075
      \\[1ex]
    \end{tabular}
  }
  \caption{\emph{Geometrical parameters of models.}
    \protect\subref{tab:cell:outline}: Mapping from geometrical positions in
    Figure~\ref{fig:cell} to indexed geometrical parameters. Inflection points
    and lengths are varied among different models, depending on counter-charge
    positions, helix conformation and gating pore size. Positions 
    in $r$ (radial) and $z$ (axial) coordinates are marked by a colored
    point and an associated number. All corners are rounded with curvature
    radius of 0.15~nm. Points 1,2,3,4, and 6 define the profile of the lipid
    and protein dielectrics, from the outermost end of the lipid domain (1)
    to the face of the S4 cylinder (6). Point 5 marks the radial position
    assigned to counter-charges, point 7 the radial position of S4 charges.
    \protect\subref{tab:cell:params}: Coordinates (in nm) of the points defining
    membrane and protein metrics. For models with symmetrical gating canal,
    $r$ and $z$ define both the extracellular and
    intracellular surfaces. For models with an asymmetric gating
    canal, $+z$ values are used for the extracellular surface and $-z$
    for the intracellular surface. These metrics apply regardless of
    electrode positioning. The numbers in parentheses listed in the first
    column refer to the variants of gating canal geometry used in the models for
    Fig.~\ref{fig:gatpore}. See
    Figs.~\ref{fig:movie:first}--\ref{fig:movie:last} for 3
    dimensional representations of some geometries.}
  \label{tab:cell}
\end{table}

We analyze two degrees of freedom for VS charge movement: the curve on
which (triplets of) S4 charge centers are aligned can be both
translated along the helix axis and rotated about that axis. The model S4
charges thus move like parts of a solid body. Negatively charged
residues contributed by the S2 and S3 transmembrane segments of the
natural VS are modeled as point charges of -1~$\text{e}_0$ arranged on a
curve parallel to the curve on which S4 charges are centered (red
spheres in Fig.~\ref{fig:helix}); the offset from the helix axis of
the counter-charge curve is 0.466~nm larger than the radius of the
curve of the centers of the (triplets of) S4 charges (see
Table~\ref{tab:cell}). The axial and angular intervals between
counter-charges are chosen so that at most one S4 charge can exactly
line up with any of the three counter-charges (see legend of
Fig.~\ref{fig:helix}). Counter-charges are stationary in their
assigned positions.

We design the simulation to address physical
questions with computational efficiency. For instance, the electrodes
encapsulating the simulation cell serve three purposes:
\begin{enumerate}
\item The bath electrodes provide Dirichlet boundary conditions
  corresponding to a voltage clamp. Our formulation of the
  electrostatics (described in the next section) establishes these
  conditions.
\item The electrodes substitute for screening by bath ions of
  uncompensated protein charge. Screening by the ions in an aqueous
  bath is equivalent to the screening provided by charge on a metal
  foil placed in the water a distance away from the protein
  boundary. According to the Debye-Hückel theory approximation, an
  electrode distance of ≈0.8~nm corresponds to physiological
  bath ionic strength. The electrode location shown in
  Fig.~\ref{fig:cell}A thus corresponds to a bath solution in the
  low millimolar range. An alternate configuration, a simulation cell
  with the bath regions omitted and the electrodes placed directly on
  the membrane and protein boundaries (Fig.~\ref{fig:cell}C),
  establishes screening at the Onsager limit (approached at
  exceedingly large ionic strength). Results obtained with the two
  alternate electrode placements allow us to examine the possible
  range of screening effects at far smaller computational cost than
  the costs of a simulation with explicit bath ions.
\item At the surface where the membrane region meets the cell
  boundary, a set of guard electrodes forming rings around the cell
  maintain a graded potential which varies between the potentials
  applied at the inner and outer bath electrodes. These guard
  electrodes impose at the membrane edge a potential similar to the
  potential existing in a macroscopic system at a location far from
  the protein.
\end{enumerate}

Translational and rotational motions of the S4 helix are simulated by
allowing the ensemble of S4 charges to slide within the dielectric
envelope provided by the S4 matrix shown in Fig.~\ref{fig:cell}. The
protein dielectric itself does not accompany the movement of the
charges. With the S4 dielectric matrix extended far enough into the
baths, keeping the S4 dielectric stationary has negligible
electrostatic consequences because the dielectric of the model is
uniform within the S4 matrix. Simulating S4 motion this way reduces
computational effort of model exploration by several orders of
magnitude.

\begin{figure}
  \centering
  \mbox{
    \includegraphics[width=0.5\columnwidth]{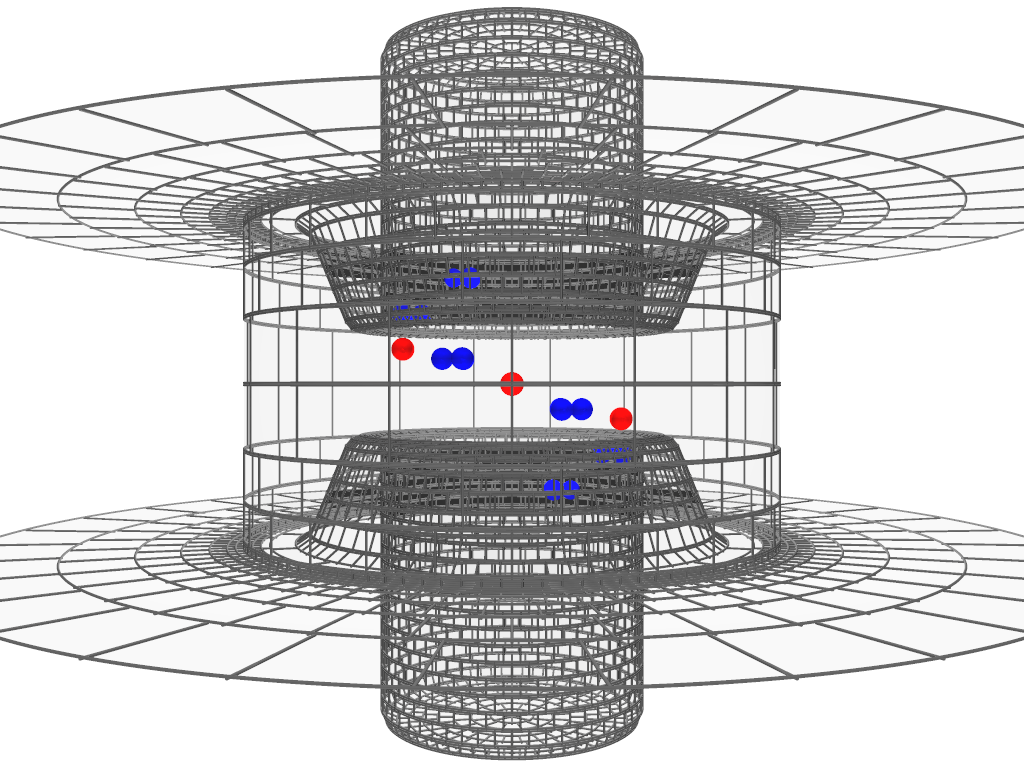}
    \includegraphics[width=0.5\columnwidth]{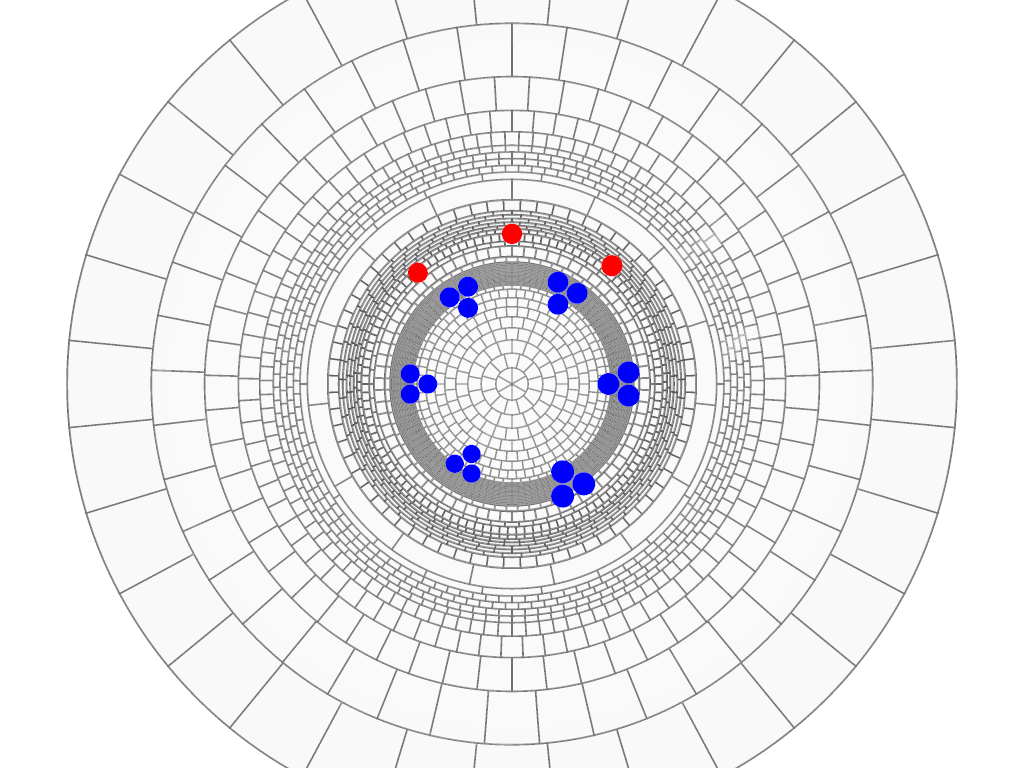}
    }
    \caption{\emph{VS charge positions and dielectric boundary
        surfaces in a \emph{α}-helical model.}  \emph{Blue symbols}: S4
      charges, each represented as three point charges of 1/3~$\text{e}_0$;
      \emph{red symbols}: counter-charges in the S2 and S3 segments,
      each represented as a single point charge of -1~$\text{e}_0$. The
      dielectric boundary surface is divided into curved tiles whose
      magnitudes are varied depending on their distance from point
      charges and local curvature.  This is the surface grid used in
      solving the induced-charge calculation.  This figure is drawn
      using perspective-enhancing features. For the range of motion of
      the S4 in this model, see the animated
      Animation~\ref{movie:1-2-3-8-0-0}.}
  \label{fig:helix}
\end{figure}

A simulation setup as shown in Fig.~\ref{fig:cell} tests a single VS
that is not interacting with other parts of the channel protein, in
particular the parts mediating the gating of the conductive pore. This
model VS thus operates as an `idling' transducer of electrical
force. The forces acting between a VS and other parts of the natural
channel are not known. The reactions of model voltage-sensors to such
forces can be explored by simulations which include hypothesized force
fields in addition to the electrostatic force field. Characteristics
of the `load' acting on the actual VS may be inferred in this
manner. However, here we report simulations of the idle VS.

\section{Methods}

\subsection{Electrostatics}

We are concerned with electrostatic interactions among charged groups
of the VS protein, electrode charges and charges induced on sharp
dielectric boundaries. In solving the electrostatics we take advantage
of the fact that all other charges besides the point charges of the
protein are distributed on a few boundary surfaces rather than
distributed throughout a volume. The primary task of calculating
electrostatic interactions consists in determining the charge
distributions on the electrode and dielectric boundaries,
distributions which are initially unknown for a given configuration of
protein charges and applied voltage.

\subsubsection{Computation of unknown charges}

\Citet{boda:2006} have described and tested a boundary element method
(Induced Charge Calculation, ICC) for computing the charge
distribution on the dielectric boundaries of a system consisting of
point charges and linear isotropic dielectrics with sharp
boundaries. We include as additional electrostatic element the
`electrode'. This element is an infinitesimally thin conductive foil
charged to a prescribed electric potential using an external
source. In our system, this surface is also a dielectric boundary
between the simulation cell and the dielectric surrounding the cell
(e.g., vacuum). Thus the charge of an electrode is in part induced
charge (as in \citeauthor{boda:2006}) and source charge (to fix the
prescribed potential). In the following we show how the total
electrode charge can be calculated without needing to separately
calculate its parts. The electrode charge calculation complements the
induced charge calculation, making it possible to calculate all the
initially unknown charges in the system.

The spatial density of source charge $\rho^{\text{src}}$ in our
system is composed of the point source charges
$\dbl{q}{k}{\text{src}}$ of the VS protein
located at positions $\mathbf{r}_k$ and source charge distributed on
the electrodes $\mathcal{E}$ with surface density
$\sigma^{\text{src}}\left(\mathbf{r}\in\mathcal{E}\right)$.

The polarization charge density induced at any location $\mathbf{r}$ in a dielectric is
\begin{equation}
\rho^{\text{ind}}\left(\mathbf{r}\right) =
  \frac{1 - \epsilon\left(\mathbf{r}\right)}{\epsilon\left(\mathbf{r}\right)}
      \rho^{\text{src}}\left(\mathbf{r}\right)
   -  \epsilon_0
      \frac{\nabla \epsilon\left(\mathbf{r}\right)}
       {\epsilon\left(\mathbf{r}\right)}
      \cdot \mathbf{E}\left(\mathbf{r}\right)
\label{eq:indcharge}
\end{equation}
where $\epsilon$ is the generally location-dependent dielectric
coefficient, $\epsilon_0$ the permittivity of the vacuum, and
$\mathbf{E}$ the electric field strength produced by all
(source and induced) charges in the system. This relation follows from
Poisson's equation (including polarization) and the constitutive
relation describing polarization in a linear, isotropic dielectric
\citep{boda:2006}. The first term on the right-hand side of
Eq.~\ref{eq:indcharge} describes the charge induced on the surfaces of
the volume element at $\mathbf{r}$ if the element contains source
charge. The second term describes charge induced in the volume element
by the electric field if the dielectric coefficient at $\mathbf{r}$
has a non-zero gradient.

For our charge calculations, we combine collocated source and induced charges into an effective charge for computing the field and potential. The effective charge density associated with a known source charge density $\rho^{\text{src}}\left(\mathbf{r}\right)$ embedded in a dielectric (described by a locally uniform $\epsilon\left(\mathbf{r}\right)$) is:
\begin{equation}
\rho^{\text{eff}}\left(\mathbf{r}\right) = \frac{\rho^{\text{src}}\left(\mathbf{r}\right)}{\epsilon\left(\mathbf{r}\right)}
\label{eq:effcharge}
\end{equation}
The effective charge density of an electrode includes both
contributions to induced charge described by Eq.~\ref{eq:indcharge},
as well as the source charge.

The dielectric boundaries $\mathcal{B}$ inside the simulation cell
(marked in \emph{black} in Fig.~\ref{fig:cell}) do not carry source
charge. However, the electric field in the simulation cell induces the
charge density $\sigma^{\text{ind}}\left(\mathbf{r}\right)$ at
locations $\mathbf{r} \in \mathcal{B}$. This induced charge density is
initially unknown.

The field strength $\mathbf{E}$ and potential $V$ in our system are produced by the superposition of the fields and potentials of the source and induced charges:
\begin{equation}
\begin{split}
\mathbf{E}\left(\mathbf{r}\right) = &
   \frac{1}{4 \pi \epsilon_0} \sum_k \dbl{q}{k}{\text{eff}}
     \frac{\mathbf{r}-\mathbf{r}_k}{|\mathbf{r}-\mathbf{r}_k|^3} \\
 & + \frac{1}{4 \pi \epsilon_0} \int_{\mathcal{B}}\!\sigma^{\text{ind}}
     \left(\mathbf{r}'\right)
     \frac{\mathbf{r}-\mathbf{r}'}{|\mathbf{r}-\mathbf{r}'|^3}
     \;da'
 + \frac{1}{4 \pi \epsilon_0} \int_{\mathcal{E}}\!\sigma^{\text{eff}}\left(\mathbf{r            }'\right)
     \frac{\mathbf{r}-\mathbf{r}'}{|\mathbf{r}-\mathbf{r}'|^3}
     \;da'
\end{split}
\label{eq:field}
\end{equation}

\begin{equation}
\begin{split}
V\left(\mathbf{r}\right) = &
   \frac{1}{4 \pi \epsilon_0} \sum_k \dbl{q}{k}{\text{eff}}
     \frac{1}{|\mathbf{r}-\mathbf{r}_k|} \\
 & + \frac{1}{4 \pi \epsilon_0} \int_{\mathcal{B}}\!\sigma^{\text{ind}}
     \left(\mathbf{r}'\right)
     \frac{1}{|\mathbf{r}-\mathbf{r}'|}\;da'
 + \frac{1}{4 \pi \epsilon_0} \int_{\mathcal{E}}\!\sigma^{\text{eff}}\left(\mathbf{r            }'\right)
     \frac{1}{|\mathbf{r}-\mathbf{r}'|}\;da'
\end{split}
\label{eq:potential}
\end{equation}
where $da'$ is the area of the surface element at location $\mathbf{r}'$.

The unknown induced surface charge density at the dielectric boundary is related to the field strength by \citep{boda:2006}:
\begin{equation}
  \sigma^{\text{ind}}\left(\mathbf{r}\right) = 
  - \frac{\Delta \epsilon \left(\mathbf{r}\right)}
         {\bar{\epsilon}\left(\mathbf{r}\right)}
  \,\epsilon_0
  \,\mathbf{n}\left(\mathbf{r}\right) \cdot
  \mathbf{E}\left(\mathbf{r}\right) 
\label{eq:chargeB}
\end{equation}
where $\mathbf{r}$ is any location on the surface $\mathcal{B}$,
$\Delta \epsilon \left(\mathbf{r}\right)$ is the change of dielectric
coefficient in the normal direction
$\mathbf{n}\left(\mathbf{r}\right)$ of the dielectric boundary, and
$\bar{\epsilon}\left(\mathbf{r}\right)$ is the mean of the dielectric
coefficients at the boundary location.

The potential at any location $\mathbf{r}$ on the electrode surfaces $\mathcal{E}$ has a value $V^{\mathrm{VC}}\left(\mathbf{r}\right)$ imposed by the voltage clamp:
\begin{equation}
V\left(\mathbf{r}\right) = V^{\mathrm{VC}}\left(\mathbf{r}\right)
\label{eq:chargeE}
\end{equation}
Inserting the expression for the electric field strength from
Eq.~\ref{eq:field} into Eq.~\ref{eq:chargeB} and inserting the
expression for the electric potential from Eq.~\ref{eq:potential} into
Eq.~\ref{eq:chargeE} yields two integral equations in terms of both
$\sigma^{\text{ind}}\left(\mathbf{r}\in\mathcal{B}\right)$ and
$\sigma^{\text{eff}}\left(\mathbf{r}\in\mathcal{E}\right)$. The
initially unknown charge densities of the dielectric and electrode
boundaries are the joint solution of these two integral equations.

To solve the integral equations we follow the method of
\citet{boda:2006}. The surfaces $\mathcal{B}$ and $\mathcal{E}$ are
subdivided into curved surface elements. The unknown charge densities
are approximated as uniform within each surface element. The two
integral equations then become one system of linear equations in terms
of the unknown charge densities of a finite number of surface
elements. The accuracy of this approximation is verified by comparing
the numerical solution of the discretized integral equations with the
analytical result of the integral equations (which is possible for
certain geometries) or by verifying Gauss's theorem for any specific
geometry of interest. Examples for the test by Gauss's theorem are
included in Supplementary Materials. This test based on Gauss's theorem
was applied for each specific simulation.

The inhomogeneity of surface charge in our simulation is greatest
where point charges of the VS protein are close to a dielectric
boundary or electrode. For computational efficiency, we vary the size
of the surface subdivisions depending on the distance from the point
charges. A typical surface grid (comprising ≈6000 surface
elements) is shown in Fig.~\ref{fig:helix} for the dielectric surfaces
in the simulation cell. The electrode surfaces (not included in
Fig.~\ref{fig:helix}) are subdivided into ≈1200 relatively
large elements because of their distance from point charges.

The computation of the unknown charges of the surface elements
involves solving a linear equation system in terms of as many unknowns
$\left(N\right)$ as there are surface elements. The coefficient matrix
of this system is dense, therefore LU-decomposition time increases
with $\mathcal{O}\left(N^3\right)$. This computational disadvantage is
greatly alleviated by the fact that $LU$-decomposition of the
coefficient matrix needs to be done only once for a given combination
of surface geometry and dielectric coefficients. When the S4 charges
are moved, solutions to this system of equations are obtained by
back-substitution using the same LU-decomposed matrix. That
$\mathcal{O}\left(N^2\right)$ operation, back-substitution, is
required for each sampled configuration of VS charges, but not for
each applied voltage tested, as described later.

\subsubsection{Computation of charge displacement and electrostatic energy}

When the charges of the VS change position, the electric flux toward
one bath electrode generally increases by the same amount as the
electric flux decreases toward the other bath electrode. To maintain a
constant voltage between the two electrodes, charge has to be moved
externally between the electrodes. This charge is the experimentally
measured displaced gating charge. In principle, one can measure
displaced charge in a simulation by monitoring electric flux across a
surface surrounding a bath electrode. A more efficient method is
provided by the Ramo-Shockley (RS) theorem
\citep{shockley:1938,ramo:1939}; for an application to ion channels
see \citep{nonner:2004:rs}). The RS theorem lays the groundwork as
well for an efficient method of computing the electrostatic energy of
VS configurations when the applied voltage is varied \citep{he:2001}. The
RS theorem is applicable to systems containing linear dielectrics.

The RS theorem can be formulated for the configuration of electrode
potentials in this study. We apply equal and opposite potentials
$V_{\mathrm{m}}/2$ and $-V_{\mathrm{m}}/2$ to the internal and
external bath electrodes to create a membrane voltage $V_{\mathrm{m}}$
(defined as internal minus external potential). This method of
applying a membrane potential maintains global electroneutrality of
the simulation system. We determine the displaced charge in a
simulation in two steps:
\begin{enumerate}
\item Set all point charges to zero and apply +1/2 volt at the
  internal and -1/2 volt at the external bath electrode. Solve for the
  unknown electrode and induced boundary charges. From the electric
  potential $V_0\left(\mathbf{r}\right)$ produced by these charges, an
  `electrical distance' can be computed as the ratio
  $\left[V_0\left(\mathbf{r}\right)\right]/\left[1\text{ volt}\right]$
  for any geometrical location $\mathbf{r}$ in the simulation cell.
\item For a simulation run with the actual point charges $q_k$ present
  and actual potentials $V_{\mathrm{m}}/2$ and $-V_{\mathrm{m}}/2$
  applied at the electrodes, determine the displaced charge $Q_k$ from the
  relation:
\begin{equation}
Q_k = q_k \frac{V_0\left(\mathbf{r}_k\right)}{1\text{ volt}} 
\label{eq:rs}
\end{equation}
with the electrical distance computed as described in step 1. When several point charges are in the simulation, total displaced charge is the algebraic sum of the displaced charges defined by Eq.~\ref{eq:rs} for each point charge: 
\begin{equation}
Q = \sum_k Q_k
\label{eq:rsall}
\end{equation}

\end{enumerate}
Step 1 is executed once for the chosen configuration of electrodes and dielectrics in the simulation (including dielectric coefficients). Step 2 is executed once for each varied configuration (or number) of point charges for which displaced charge is determined. Since the displaced charge determined in step 2 is invariant with respect to the potentials applied at the electrodes, it does not need be re-calculated for the same point-charge configuration when electrode potential is varied.

Note that the electrical distance obtained with the equal and opposite electrode potentials varies between -1/2 and +1/2 as the point charge is moved from the internal to the external bath electrode. The electrical distance zero corresponds to geometrical locations where the point charge induces equal amounts of charge on either bath electrode.

The RS theorem also makes it possible to compute the electrostatic energy in
an efficient two-step procedure \citep{he:2001}:
\begin{enumerate}
\item In the full simulation system (including the point charges),
  impose the potential $V_{\mathcal{E}}=0$ on all
  electrodes. Calculate the electrostatic interaction energy $W_1$ in
  the system with the point charges $q_k$ at positions $\mathbf{r}_k$
  from the relation:
\begin{equation}
W_1 = \frac{1}{2} \sum_k q_k V^{V_{\mathcal{E}}=0} \left(\mathbf{r}_k\right) 
\label{eq:nullenergy}
\end{equation}
where $V$ is computed with Eq.~\ref{eq:potential}.

\item Calculate the displaced charge $Q$ corresponding to the point
  charges and their positions using Eq.~\ref{eq:rsall}. For the
  imposed voltage $V_{\mathrm{m}}$, calculate:
\begin{equation}
W_2 = -QV_{\mathrm{m}}
\end{equation}

\end{enumerate}
Step 1 of this procedure is executed once for each sampled location of the point charges. Step 2 is executed repeatedly for each applied voltage that is tested. The electrostatic configurational energy in the simulation with the applied voltage $V_{\mathrm{m}}$ is the algebraic sum of the energies calculated in steps 1 and 2:
\begin{equation}
W = \frac{1}{2} \sum_k q_k V^{V_{\mathcal{E}}=0} \left(\mathbf{r}_k\right) - QV_{\mathrm{m}}
\label{eq:ework}
\end{equation}
Using the RS theorem in these secondary calculations
greatly increases computational efficiency of simulations. This is the
basis for our systematic explorations of models.

The calculation of displaced charge and electrostatic energy via
Eqs.~\ref{eq:rsall} and \ref{eq:ework} are independently verified by
computing the electrostatic work through a second route. We
alternatively compute the work from the path integral of the electric
force acting on the charges $q_k$ of the VS as those charges move from
$\mathbf{r}_k'$ to $\mathbf{r}_k''$:
\begin{equation}
  \label{eq:ework2}
  \Delta W = \sum_k q_k
    \int_{\mathbf{r}_k'}^{\mathbf{r}_k''}
    \hskip-1.5ex\mathbf{E}(\mathbf{r}_k)
    \cdot d\mathbf{r}_k
\end{equation}
Here, the electric field is the field of all charges in the system,
including the electrode charges provided by the external source that
maintains the electrode potentials as VS charges move along their
paths. An example of this control (showing consistency) is given in
Supplementary Materials (Fig.~\ref{fig:Ebypath}).

\subsection{Statistical mechanics}

Displaced gating charge is experimentally measured from ensembles of
channels and thus is an ensemble average. Our electrostatic
calculations yield both the displaced charge and the electrostatic
part of the configurational energy for a given configuration of a
simulated VS model. We consider whole-body movements of S4 charge in
two degrees of freedom: translation along the S4 axis and rotation
about that axis. Our computational method is efficient enough to allow
systematic sampling of this configuration space. We represent each
dimension by 51 equally spaced grid nodes and compute the electrostatic
energy for the 2601 nodes of the two-dimensional space.

The energy samples define a canonical partition function describing
the consequences of the electrostatics on the distribution of an
ensemble in the discretized configuration space:
\begin{equation}
\mathcal{Q} = \sum_{i,j} e^{-W_{ij}/k_BT}
\label{eq:partitionfct}
\end{equation}
where $i$ and $j$ are the indices of the rotational and translational
discrete positions; $W_{ij}$ is the electrostatic configurational
energy of the voltage sensor at translational position $i$ and
rotational position $j$; $k_B$ is the Boltzmann constant; and $T =
298.15$~K is the absolute temperature. The sampled rotational range is
360\textdegree, and a typical translational range is -1.925~nm to
+1.925~nm relative to the central position of the S4 charges in
α-helical models. Configurations near the ends of the chosen
translational range have very small probabilities due to their large
electrostatic energy.

The probability of a VS configuration is then:
\begin{equation}
  \label{eq:dist}
  P_{ij} = \frac{1}{\mathcal{Q}}e^{-W_{ij}/k_BT}
\end{equation}
and the expectation value of a random variable $X$ is:
\begin{equation}
\langle X\rangle = \sum_{i,j} X_{ij}P_{ij} 
= \frac{1}{\mathcal{Q}}\sum_{i,j} X_{ij} e^{-W_{ij}/k_BT}
\label{eq:randvar}
\end{equation}
One example of a random variable of interest is the charge displaced
at an applied voltage. By applying Eq.\ref{eq:dist}, the
charge/voltage relation for the model can be predicted for direct
comparison with experimental observation of that relation.

We also determine expectation values of random variables over the
rotational degree of freedom for a particular translational position
$i$:
\begin{equation}
\langle X_i\rangle = \sum_{j} X_{ij}P_{ij} 
= \frac{1}{\sum_{j} e^{-W_{ij}/k_BT}}\sum_{j} X_{ij} e^{-W_{ij}/k_BT}
\label{eq:randvarrot}
\end{equation}

The (Helmholtz) free energy of our canonical simulation system is given by the statistical-mechanical relation:
\begin{equation}
A = -kT \ln {\mathcal{Q}}
\label{eq:helmholtz}
\end{equation}

\subsection{Online Supplemental Materials}

Figs.~\ref{fig:gausstest} and \ref{fig:Ebypath} provide controls for the numerical accuracy in solving the model. One tests the computation of charge induced by the VS charges on the boundary of the protein dielectric and is based on Gauss's law. The other tests the computation of electrostatic energy by an independent method (by integrating force along a given S4 path).

Figs.~\ref{fig:movie:first} through \ref{fig:movie:last} and the
associated animations (Animations~\ref{movie:first}--\ref{movie:last}
respectively) illustrate VS geometry or movement in four simulations
presented in this paper. They show the mean position(s) of VS charges
superimposed to the distribution of charge density in the microscopic
system. In the animations, voltage is changed in a ramp from -100 to
+100~mV.

\section{Results and Discussion}

The relation between the displaced gating charge and the applied
membrane potential in a sliding-helix model of an individual VS domain
is shown in Fig.~\ref{fig:qvsbothhelices}A. The solid line represents
the computed relation. Symbols reproduce the relation experimentally
observed in \emph{ShakerB} potassium channels
\citep[Fig.~2A]{seoh:1996} --- the experimental charge per channel was
divided by the number of channel monomers, 4. Three observations can
be made by comparing the two relations: (1) the total amounts of
charge that can be moved by large changes of voltage are similar,
≈3 elementary charges per VS domain; (2) the slopes of the
two relations are similar, but not equal; (3) a shift along the
voltage axis is needed to align the midpoint of the computed relation
to the midpoint of the experimental relation (dotted line).

\begin{figure}
  \centering
  \includegraphics{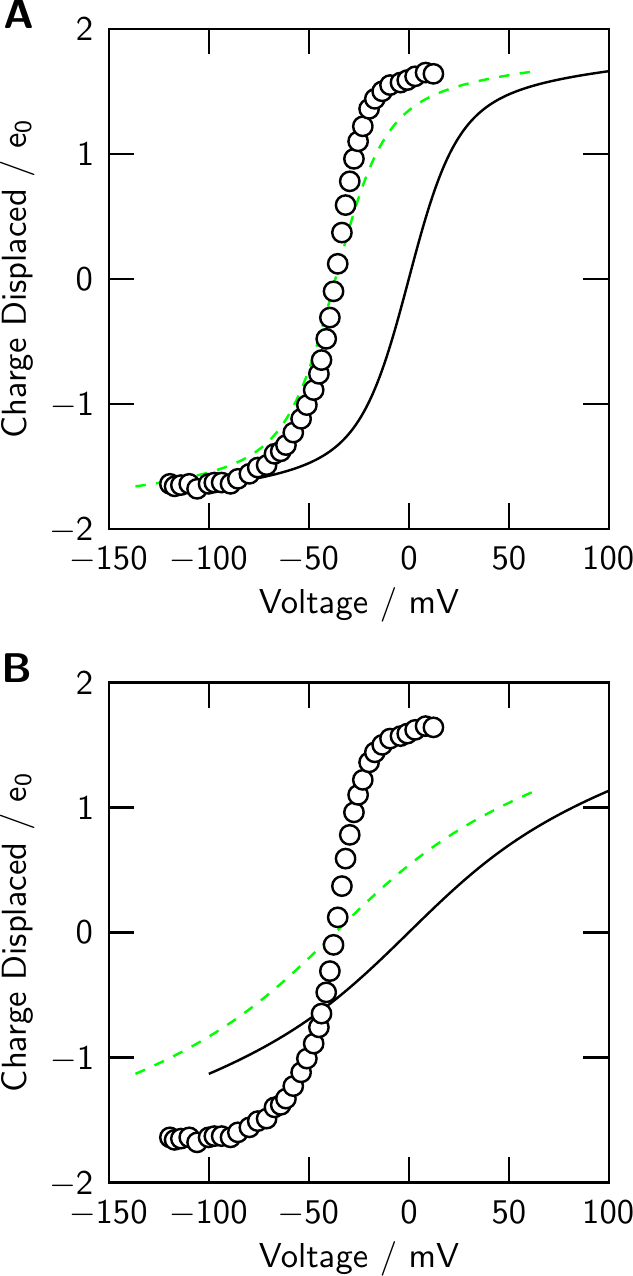}
  \caption{\emph{Theoretical charge/voltage relations compared to
      experiment.} (A) α-helical S4; (B) $3_{10}$-helical
    S4. Computed relations are shown as solid lines. An experimental
    relation for \emph{Shaker} $\text{K}^+$ channels \citep{seoh:1996} is
    shown as symbols. Dashed lines: computed curves shifted to match
    the midpoint of the experimental curve. Simulations of the
    `standard' models are described in section `Model and boundary
    conditions'. The counter-charge spacing is set to 2/3 of the
    (translational and rotational) interval of the S4 charges, for
    both the α- and the $3_{10}$-helical S4 models.}
  \label{fig:qvsbothhelices}
\end{figure}

The VS charges in the simulation for Fig.~\ref{fig:qvsbothhelices}A
are arranged according to an α-helical geometry
(Figs.~\ref{fig:helix} \& \ref{fig:movie:standard}). An analogous
simulation in which the VS charges conform to a $3_{10}$-helical
geometry (Fig.~\ref{fig:movie:310}) yields a different result
(Fig.~\ref{fig:qvsbothhelices}B). With regard to the experimental
charge/voltage relation, the $3_{10}$ model yields less total
charge movement over the tested voltage range and a smaller maximum
slope.

The model parameters giving the charge/voltage relation in
Fig.~\ref{fig:qvsbothhelices}A are the initial reasonable
choices/guesses made in this study as a basis for comparison, except
for the spacing of counter-charges which was chosen from three
variations made of the α-helical S4 model. The spacing used
prevents S4 charges from simultaneously aligning with more than one
counter-charge within the average range of travel
(Fig.~\ref{fig:qvsbothhelices}, legend). The counter-charges in the
$3_{10}$-helical model of Fig.~\ref{fig:qvsbothhelices}B were arranged
using the same rule of spacing between counter-charges. No attempt is
made at this stage to find counter-charge positions giving potentially
more realistic charge/voltage relations for the $3_{10}$-helical
model.

In the following, we use the models of Fig.~\ref{fig:qvsbothhelices}
as `standard' models that are altered to explore sensitivity to
variation of physical or geometrical parameters. The primary questions
we address are: (1) Which parameters are more (or less) important for
gating charge displacement? (2) In which ways are these parameters
important for a sliding helix mechanism? Our simulations with varied
parameters reveal model designs that predict experimental relations
between charge displaced and voltage more closely than does the
standard model used for Fig.~\ref{fig:qvsbothhelices}A. However to
make comparisons of experimental and simulated relations, at least
three caveats need to be considered:
\begin{enumerate}
\item Experimental gating charge is measured using a `$P/n$' protocol
  \citep{armstrong:1977} by which linear charge displacement is
  eliminated from the record to uncover non-linear charge displacement
  presumably associated with VS motion. If a $P/n$ protocol measures
  `linear' charge displacement between voltages at which VS charge
  displacements are not equal, the protocol eliminates displacement
  associated with the VS itself (scaled by the factor $n$). The
  theoretical charge/voltage relations shown here comprise \emph{all}
  displaced charge. For an accurate comparison to experiment, these
  relations would have to be subjected to the specific $P/n$
  subtraction procedure used in the experiment. The unsubtracted
  relations that we present contain the information needed to decide
  where a $P/n$ subtraction procedure would eliminate actual gating
  charge (e.g., the simulation results of
  Fig.~\ref{fig:qvsbothhelices}B).

\item Theoretical charge displacement is associated with a known
  geometrical position of the VS charges. Experimental charge
  displacement is not referenced to a known geometrical position. To
  compare with theory, we assign the experimental charge displacement
  a reference midway between zero and the charge measured at the
  largest tested depolarization. In interpreting experiments, that is
  an arbitrary choice to be made for each version of natural or
  engineered channel. We represent the simulated charge displacement
  as relative to the electrical distance 0 as defined in Methods. In
  the case of the models for Fig.~\ref{fig:qvsbothhelices}, this
  electrical distance coincides with the geometrical translation $z=0$
  because of the symmetry of the dielectrics.

\item The VS domain in the simulation operates under different
  conditions from the VS domains in a full channel. The model
  transduces voltage changes solely into changes of position of the
  VS. The VS domains integrated into a channel also transduce voltage
  changes into action on other domains controlling ion flow in the
  central pore. A shift along the voltage axis (as suggested by
  Fig.~\ref{fig:qvsbothhelices}) is one possible way in which a `load'
  may be reflected in the charge/voltage relation of the VS.
\end{enumerate}

The simulations reported in Fig.~\ref{fig:qvsbothhelices} indicate
that voltage sensing by a sliding helix is robust from an engineering
point of view. Either the α- or $3_{10}$-helical S4 structure can
produce voltage-dependent charge displacement, even though the
structures lead to distinct configurations of the charges. The two
forms of helix generate different voltage sensitivities in the tested
models -- but neither form produces catastrophic failure.

The simulated VS models we use have been reduced to a simple geometry
and to the electrostatic interactions of point charges and linear
dielectrics. Since those models can produce more-or-less realistic
behavior, important insights into the VS system may be possible by the
rigorous analysis of a few crucial components in a reduced model. A
reduced model allows physical features to be changed in an orthogonal
and systematic manner. The variations presented in this paper all
start from the two `standard' models introduced in this section. A key
variable for analyzing change is the configurational energy arising
from the electrostatics.

\subsection{Energetics of sliding-helix models}

The mobile charges of the VS model are within the electric field of
the charges on the electrodes, of the stationary counter-charges, and
of the charges induced in the dielectrics. The charges of the S4 helix
are modeled as parts of a solid body. We consider two degrees of
freedom of solid-body motion: translation along the S4 axis and
rotation about that axis. Energy is computed on a grid composed over
this configuration space (Eq.~\ref{eq:ework}). From the electrostatic
energy map, a partition function is constructed
(Eq.~\ref{eq:partitionfct}), and from the partition function,
statistics of random variables (Eq.~\ref{eq:randvar}) are
predicted. The energy map thus defines the expectations of observables
such as the gating charge displacement
(Fig.~\ref{fig:qvsbothhelices}). The random variables computed on this
basis do not involve approximations based on interpreting energy maps
in terms of equilibrium and transition states \citep{lecar:2003}.

\begin{figure}
  \vskip-2\baselineskip
  \centering
  \includegraphics{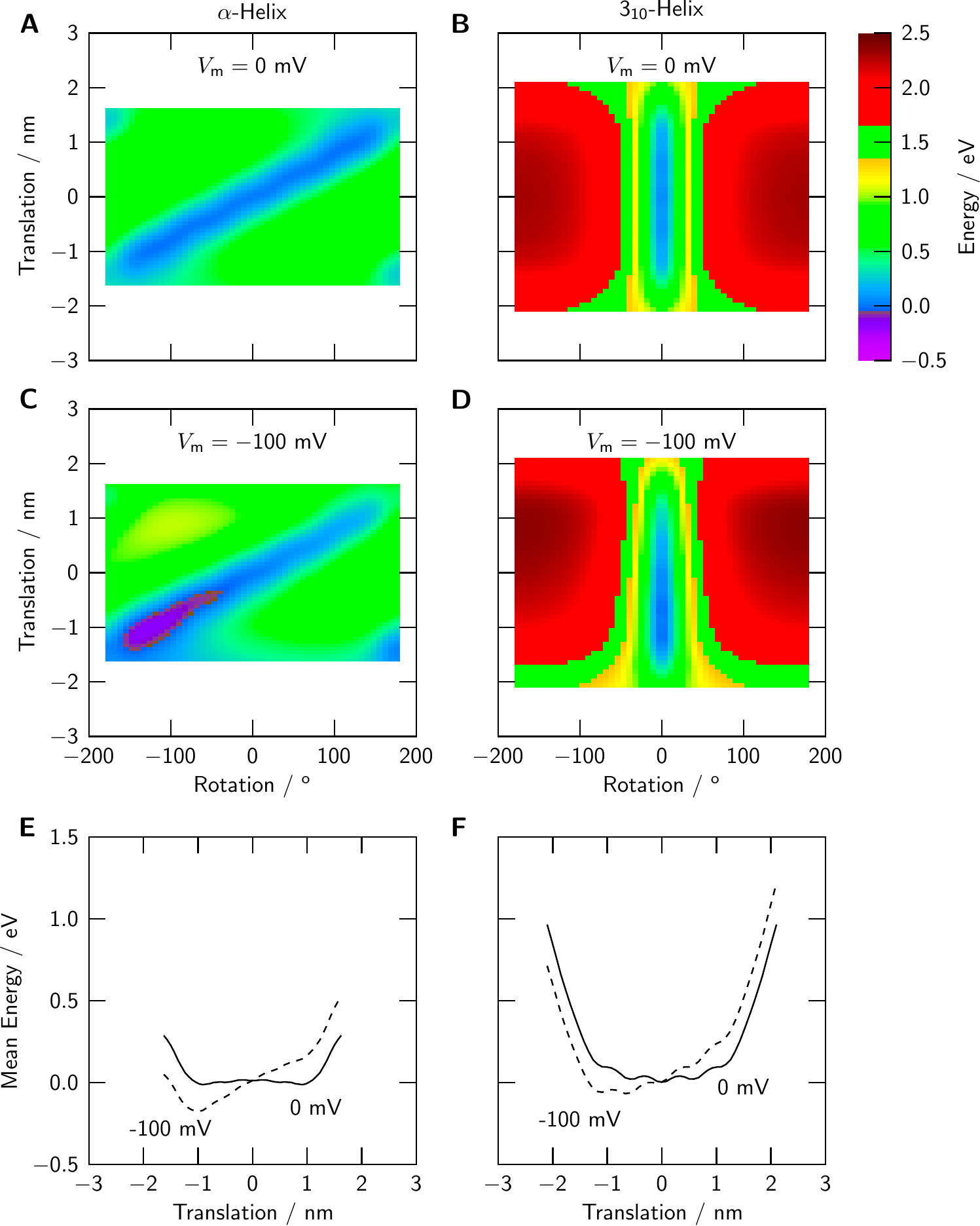}
  \caption{\emph{Energetics of voltage sensing.} (A, C, E) α-helical S4; (B, D, F) $3_{10}$-helical S4. Pseudo-color maps: electrostatic energy for applied voltages 0~mV (A, B) and -100~mV (C, D). (E, F): statistical-mechanical expectation of energy over the rotational degree of freedom versus translation; solid line: 0~mV; dashed line: -100~mV.}
  \label{fig:Ebothhelices}
\end{figure}

Pseudo-color maps of energy in the two dimensions computed for applied
voltages of 0 and -100~mV are shown in
Fig.~\ref{fig:Ebothhelices}. The translational and rotational origin
corresponds to the central S4 position shown in Fig.~\ref{fig:helix}
for the α-helix. Energy is presented as relative to that at the
origin. The symmetry of these models implies that the map
for +100~mV (not shown) is a mirror image of the map for -100~mV.

For a membrane potential of 0~mV, the energy maps of both helical
models reveal a trough bounded on all sides by regions of
substantially higher energy. The energy trough runs in the direction
of proportional translation and rotation for the α-helical S4 model,
but it runs in the direction of simple translation for the
$3_{10}$-helical S4 model. The trough in the map of each model follows
the counter-charge arrangement --- an arrangement chosen for each
model to allow periodic interactions during S4 motion of the S4
charges with the counter-charges. In the α-helical model, the S4
charges and counter-charges are aligned on parallel super-helices,
whereas in the $3_{10}$-helical model the S4 charges and
counter-charges follow straight lines parallel to the helical
axis. Lowest electrostatic energies of the α-helical S4 trace the path
of a screw, whereas energies of the $3_{10}$-helical S4 trace the path
of a piston.

The regions outside the energy trough for the $3_{10}$-helix have energies about three times as large as those of the α-helix. The electrostatic confinement of the $3_{10}$-helix is stronger than that of the α-helix. The strength of confinement correlates inversely with the separation of charges in the two  geometries. The cluster of S4 charges and counter-charges is more spread out in space in the α-helical than in the $3_{10}$-helical geometry due to the angular separations of charges in the α-helical S4.

The electrostatic energy trough tends to anchor the sliding helix in a
transmembrane configuration. The S4 charges that dwell in a given
configuration in the region of small polarizability are balanced in
the model by counter-charges located in that region. This balance is
maintained over the range of S4 travel where equivalent amounts of S4
charge and counter-charge overlap in the region of weak dielectric
(see Fig.~\ref{fig:movie:standard} and
Animation~\ref{movie:1-2-3-8-0-0}). A second essential element of
balance concerns the transit of S4 charges between the less
polarizable gating canal region, and the more strongly polarizable
vestibule and bath regions. Any energy change associated with the
transit of an S4 charge on one side is balanced on the other side by
the opposite transit of an S4 charge. On the other hand, the energy
trough generated by the electrostatics of the models is too shallow by
itself to ensure long-term stability of the S4
configuration. Stabilizing interactions beyond those included in the
model (such as interactions provided by the linkers to adjacent
transmembrane segments) are most likely necessary for long-term
stability.

To inspect the energetics more closely, we construct a
one-dimensional energy profile for S4 translation by computing for
each translational position the expectation of the electrostatic
energy (in the rotational degree of freedom using the rotational
partition function, Eq.~\ref{eq:randvarrot}). We refer to this kind of
energy profile as a `translational energy profile' for
short. Fig.~\ref{fig:Ebothhelices}E,F show the translational energy
profiles for two applied voltages: 0~mV (\emph{solid} lines) and -100
mV (\emph{dashed} lines). The profiles at 0~mV are quite uniform over
the translational extent of the energy trough. At a membrane voltage
of -100~mV, the energy profiles are tilted in favor of more
intracellular positions. A well-defined energy minimum is found at a
position about -1~nm inward from the central position of the
α-helical S4, whereas a broad minimum spread between -0.7 and
-1.2~nm is found with the $3_{10}$-helical S4.

\begin{figure}
  \centering
  \includegraphics{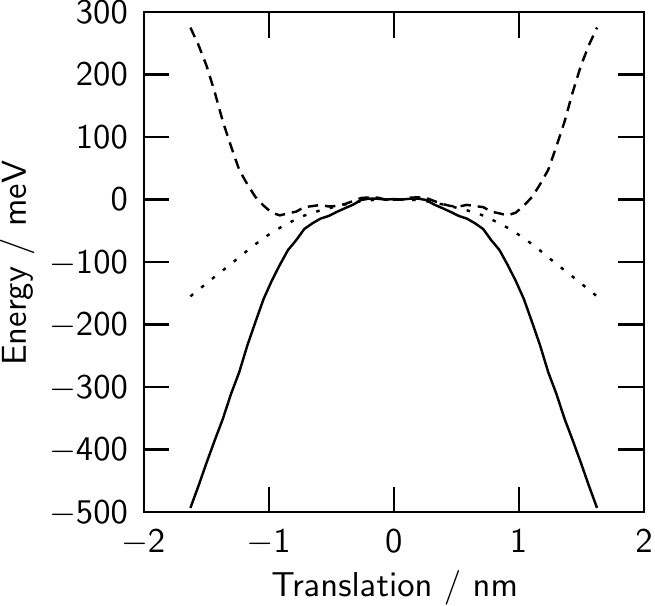}
  \caption{\emph{Energetic consequences of the counter-charges.} Translational energy profile of the α-helical model of Fig.~\ref{fig:qvsbothhelices}A with counter-charges present (dashed line) or deleted (solid line); $\epsilon_{\mathrm{p}}=4$. The dotted curve is computed with counter-charges deleted and $\epsilon_{\mathrm{p}}=16$}
  \label{fig:nocounter}
\end{figure}

The energy profiles in Fig.~\ref{fig:Ebothhelices} do not resemble the
profile of an ion embedded in a lipid membrane --- the latter profile
has a high barrier in the center of the weak dielectric
\citep{parsegian:1969,neumcke:1969}. Instead, in the models the
charged section of the S4 helix can travel with an almost level energy
over the range of translation where S4 charges overlap with stationary
counter-charges. To examine the contribution of the counter-charges to
this result, we re-compute the energy profile for the α-helical
S4 with the counter-charges deleted from the model
(Fig.~\ref{fig:nocounter}). Deletion of the counter-charges converts
the energy trough seen with the standard model (\emph{dashed line})
into a broad barrier (\emph{solid line}). That barrier is reduced but
not inverted by increasing the VS dielectric coefficient
$\epsilon_{\mathrm{p}}$ from 4 to 16 (\emph{dotted line}). For
comparison, a sliding-helix model without counter-charges using
$\epsilon_{\mathrm{p}}=10$ has been analyzed previously by
\citet{grabe:2004}.

Over the range of translation where the counter-charges produce an
energy trough (\emph{long-dashed line}), the deletion of the
counter-charges (\emph{solid line}) has a rather small effect (note
that we plot energy as relative to the central position). Movement of
the S4 helix encounters little energy variation in this region of
translation because the amount of S4 charge present in the domain of
weak dielectric does not vary: as S4 charge enters on one side, S4
charge leaves on the other side. The energetics are not favorable for
these S4 positions (as the charges are not balanced), but they are
rather uniformly unfavorable until S4 translation exceeds a distance
of about 1~nm from its center position. If the travel of the S4 helix
is restricted to $\pm$1~nm in a biological channel by means other
than the counter-charges deleted here, then even an S4 without
balancing counter-charges could perform as a voltage sensor.

The variations of energy are small over the traveled range of
translation in Fig.~\ref{fig:Ebothhelices}E,F. The restriction in the
model that the S4 domain and its charges must move as a single solid
body might be expected to `synchronize' periodic interactions among
charges and counter-charges, leading the energetics to express several
distinct barriers and wells. The chosen spacing of the
counter-charges, however, is enough to prevent the emergence of such a
pattern. Additional degrees of freedom are thus not a prerequisite for
smooth S4 travel (examples of such degrees of freedom are the possible
flexibility of the individual charge-bearing S4 residues or changes in
configuration of the helix between the α and $3_{10}$ forms,
\citealp{long:2007,khalili-araghi:2010,bjelkmar:2009,schwaiger:2011}),

Although the energy profiles of the two helix forms are similar, the
small differences between them are sufficient to produce substantial
differences in the relations between charge displacement and voltage
(Fig.~\ref{fig:qvsbothhelices}). Simulation results with varied
electrostatic parameters presented in the following reveal a number of
ways that VS characteristics may be tuned. Because small energy
variations control VS behavior, variations in numerous other
characteristics of the channel protein might achieve such tuning.

\subsection{Voltage sensitivity}

\citet{hodgkin:1952:quant} discovered ``the extreme steepness of the
relation between ionic conductance and membrane potential''. They
attached physical significance to this steepness by supposing that the
ion flow is controlled by the distribution of a charged particle that
``may rest either on the inside or the outside of the membrane but is
present in negligible concentrations elsewhere.'' Using Boltzmann's
principle, they derived a relation between conductance and voltage in
which steepness is determined by the valency $z$ of the charged
particle. The `effective valency' $z$ has been used ever since to
empirically quantify the voltage sensitivity expressed in the
relations between voltage and ionic conductance, as well as the
relations between voltage and displaced gating charge.

The effective valency is model dependent because its physical meaning is defined through the two-state model. The physical meaning of this effective valency is not clear if any other mechanism is at work. Since we study a range of models, a more general definition of voltage sensitivity is needed. 

The essence of a voltage sensor is that it derives a change in free
energy from an external electrical field, a change which can then
drive other processes requiring an increase in free energy in other
parts of the channel, such as gating. Therefore to generalize voltage
sensitivity, we consider the relation between free energy and applied
voltage in our models. Since we simulate a canonical system, the free
energy in our case is the Helmholtz energy $A$ as defined by
Eq.~\ref{eq:helmholtz}.

\begin{figure}
  \centering
  \includegraphics{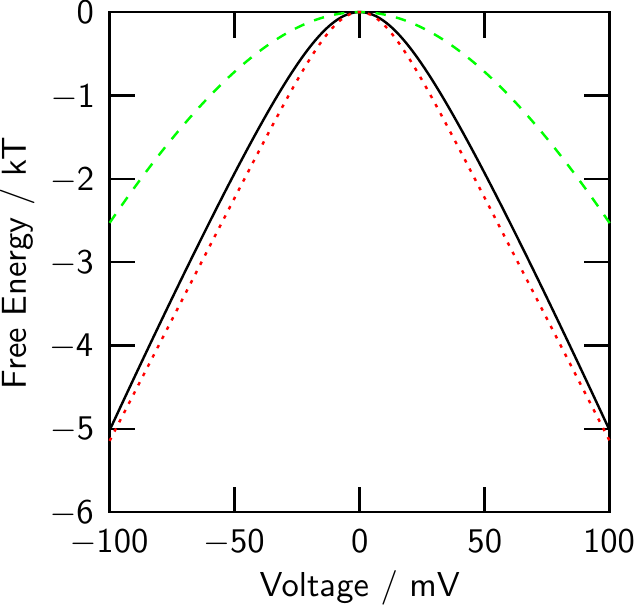}
  \caption{\emph{Voltage dependence of free energy.} Free energy
    (Eq.~\ref{eq:helmholtz}) for: α-helix model (solid line),
    $3_{10}$-helix model (long-dashed line) and two-state analytic
    Boltzmann model with $z=3$ (dotted line). All free energies are
    relative to the energy at $V=0$.}
  \label{fig:Abothhelices}
\end{figure}

Fig.~\ref{fig:Abothhelices} shows the free-energy/voltage relations
for our standard α-helical VS model (solid line), our standard
$3_{10}$-helical model (long-dashed line), and the analytically
derived relation for the \citeauthor{hodgkin:1952:quant} two-state VS
model using $z=3$ (dotted line). The plotted free energies are
relative to the free energy applying to $V=0$. In all cases,
increasing the absolute value of the voltage reduces the free energy
of the VS system, but the $3_{10}$-helical model responds less steeply
to voltage than the other models. The α-helical and two-state
models produce quite similar changes in free energy.

Applied voltage reduces the free energy of the VS system because
states of lower electrostatic energy become available. The free energy
does not follow exactly the variation of the expectation of
configurational energy (not shown), because the free energy includes
both the configurational energy and an entropy term due to the
redistribution of the voltage sensor in the configuration space. In
microscopic systems like the VS, the entropy contribution to free
energy is substantial.

It is important to note that the reduction of free energy with applied
voltage becomes steeper toward voltages where charge displacement
(Fig.~\ref{fig:qvsbothhelices}) becomes less steep or
saturates. Voltage sensitivity as defined by the slope of the
free-energy/voltage relation is then a function over the membrane
potential and not just a singular number. This sensitivity is smallest
near zero voltage where the charge displaced varies most steeply for
these models. Thus the free energy drop due to applied voltage is
small where most of the energy received by the VS is used for
re-distributing the S4 helix among possible translations. As the
re-distribution saturates, free energy approaches a linear
relationship with voltage (as does the energy) because the electrical
potential at which the S4 charges are located varies linearly with
applied voltage. Free energy thus varies the most where charge
redistribution is minimal. The asymptotic slopes of the
free-energy/voltage relation are determined by the maximal gating
charges that can be displaced in either direction from the translation
corresponding to electrical position zero.

These observations on free energy/voltage relations show that it is
not the movement itself of the S4 helix that produces the free energy
picked up by the VS. Actually, the free energy is determined by the
increase in the electrical field acting on the S4 charges that have
reached nearly saturated positions. Thus the usefulness of the VS
`muscle' is chiefly in isometric work or against a workload requiring
a large membrane voltage to overcome. The relations shown in
Fig.~\ref{fig:Abothhelices} describe the idle VS with no work done on
other parts of the channel. The change in free energy due to applied
voltage is thus an upper limit for the free energy available to other
parts of the channel.

Since the physiological voltage range is smaller than the range
simulated in Fig.~\ref{fig:Abothhelices}, the `sluggish'
redistribution of VS charge of the $3_{10}$-helical model causes a
severe limitation in the amount of work that can be delivered by this
sensor. This limitation occurs despite the fact that the electrostatic
energy trough (Fig.~\ref{fig:Ebothhelices}F) in principle allows a
range of S4 play comparable to that in the α-helical model
(Fig.~\ref{fig:Ebothhelices}E). The full work corresponding to the
total gating charge that the $3_{10}$-helical S4 can produce at
extreme voltages cannot be delivered within the physiological voltage
range because the gating charge is not rearranged within a narrow
interval of voltage.

In the simulation studies that follow, the total charge moved is a
more robust property than the swift redistribution of gating charge in
response to voltage, which is a property more sensitive to variation
in many our model parameters. Both aspects limiting voltage
sensitivity, the total gating charge and the steepness of charge
re-distribution, are readily recognized in the experimentally
observable charge voltage relation, provided that the experimental
charge per VS is determined. Therefore we will discuss parameter
variations using charge/voltage relations rather than
free-energy/voltage relations.

\subsection{Polarizability of the protein}

The dielectric coefficient $\epsilon_{\mathrm{p}}$ assigned to the VS
in our models parameterizes the polarizability of a protein, and thus
implicitly describes re-adjustments of electrons, atoms and groups of
atoms in response to S4 movement. The dielectric coefficient implies a
large number of internal degrees of freedom beyond the two degrees of
freedom in whole-body motion sampled explicitly. It is important
therefore to determine how much electrical polarization can contribute
to shaping VS characteristics. We assess the consequences of
polarization for model behavior by varying $\epsilon_{\mathrm{p}}$
over a range of values used in other theoretical studies of proteins
(reviewed by \citealp{schutz:2001}) without concerning ourselves with
the physics underlying polarization in the VS. Our computations give
consistent solutions for a system with piecewise homogeneous linear
dielectrics and include appropriate boundary conditions.

A large dielectric coefficient $\epsilon_{\mathrm{p}}$ reduces the
effective charges of the VS (Eq.~\ref{eq:effcharge}), the field
produced by the effective charges inside the protein
(Eq.~\ref{eq:field}), and thus the charge induced on the protein
boundary (Eq.~\ref{eq:chargeB}). A large $\epsilon_{\mathrm{p}}$
weakens the local interactions of S4 charges with counter-charges and
induced charges, flattening the energy landscape for S4 movement.

Using the α-helical S4 model of
Figs.~\ref{fig:qvsbothhelices}A, \ref{fig:Ebothhelices}A,C, the
translational energy profiles are computed with varied protein
dielectric coefficients $\epsilon_{\mathrm{p}}$
(Fig.~\ref{fig:epsilon}A). The profiles of the dominant energy trough
become flatter as $\epsilon_{\mathrm{p}}$ is stepwise increased from
2 to 16. Even with $\epsilon_{\mathrm{p}}=2$, translational energy
in the trough varies by no more than 0.05~eV (≈2~kT). 

As described earlier, moderate variations in energy profile
significantly alter the charge/voltage relation
(Fig.~\ref{fig:epsilon}B). The steepness of charge re-arrangement is
reduced as $\epsilon_{\mathrm{p}}$ is increased. Asymptotic behavior
changes from approximately exponential toward hyperbolic.

\begin{figure}
  \centering
  \includegraphics{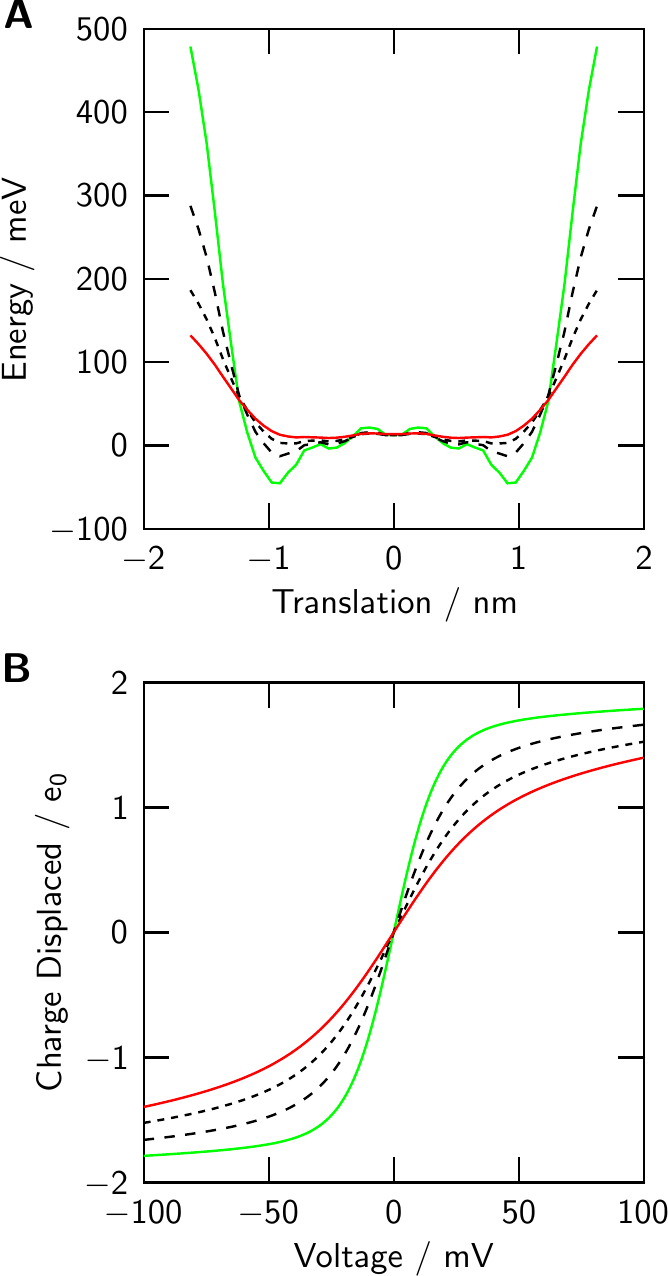}
  \caption{\emph{Effects of protein dielectric coefficient.} The
    standard α-helical model is solved for 4 different settings
    of $\epsilon_{\mathrm{p}}$: 2 (\emph{solid green line}), 4
    (\emph{short-dashed line}), 8 (\emph{long-dashed line}), and 16
    (\emph{solid red line}). A: translational energy profiles; B:
    expectation of the charge/voltage relation.}
  \label{fig:epsilon}
\end{figure}

All simulations for Fig.~\ref{fig:epsilon} use the same dielectric
geometry, protein charges and counter-charges in the
model. Nevertheless, the `effective gating charge' of the VS as
assessed by the steepness at the midpoint of the charge/voltage curve
increases as the protein dielectric coefficient is reduced. The
energetic basis for this effect of the dielectric coefficient is
evident in Fig.~\ref{fig:epsilon}A. With a dielectric coefficient of
2, there are two distinct energy minima at translations of
$\pm$1~nm. With a dielectric coefficient of 16, however, there is
very little energy variation among the positions within the energy
trough (in the absence of an applied voltage).

\enlargethispage{-2\baselineskip}
The distribution of S4 translational positions varies between two
extremes: one, a discrete distribution commonly called a `two-state
Boltzmann distribution'; the other, a continuous distribution within a
space of uniform potential energy. For a graphical representation of
these differences in distribution, see the discrete distribution of
charge in the Animation~\ref{movie:1-2-3-8-0-0} with $\epsilon=4$ and
the continuous distribution in Animation~\ref{movie:1-7-3-8-0-0} with
$\epsilon=16$. For the extremes of the discrete and continuous
distributions, the midpoint slopes of charge/voltage curves are
analytically known. The midpoint slope of the charge/voltage curve is
three times greater for the two-state case than for the uniform-energy
case \citep{neumcke:1978:gating}. Thus, the same structural charges
produce up to a three-fold varying `effective gating charge' as the
protein dielectric coefficient (and therefore the potential energy
landscape) is varied. Furthermore, the analytical asymptotes are
exponential for the two-state energy profile, but hyperbolic in
voltage for the flat energy profile.

Published theoretical studies of the VS used
$\epsilon_{\mathrm{p}}=15$ \citep{lecar:2003} or
$\epsilon_{\mathrm{p}}=10$ \citep{grabe:2004}). In our simulations,
protein dielectric coefficients from 2 to 16 have small effects on
the translational energy profile. The profile is dominated by an
energy trough that allows for the translation range where the
intrinsic counter-charges of the VS overlap with an equivalent number
of S4 charges. This trough is in fact deeper for smaller values of
$\epsilon_{\mathrm{p}}$ than for larger values
(Fig.~\ref{fig:epsilon}A). With charge balance established inside the
region of weak dielectric, charges induced on the aqueous boundaries
of the dielectric are small and their interaction with S4 charges does
not dominate the translational energy profile.

In a $3_{10}$-helical model using $\epsilon_{\mathrm{p}}=4$
(Figs.~\ref{fig:qvsbothhelices}B and \ref{fig:Ebothhelices}F), a
slightly V-shaped energy profile produces an even smaller voltage
sensitivity than that seen in the α-helical model using
$\epsilon_{\mathrm{p}}=16$ (Fig.~\ref{fig:epsilon}B). The
$3_{10}$-helical model in the presented form has difficulty producing
steep charge/voltage relations like those observed in biological
channels. On the other hand, the energetic differences determining
slope are small, so that the seemingly unfavorable characteristics of
the $3_{10}$-helical S4 might be overcome in numerous ways in
biological channels.

The simulations with varied polarizability also give information regarding the physical origin of the energy minima seen at the ends of the range of travel of the α-helical model and the small upward step in electrostatic energy seen in these positions for the $3_{10}$-helical model. It is important to note that the shown energy profiles are relative to those in the center position of these helices -- these are not the absolute energies associated with assembling the system (which are not shown). In absolute terms, electrostatic energies increase as protein dielectric coefficient is reduced. The energy wells seen in profiles associated with small polarizability reflect a smaller increase of electrostatic energy in some positions. These wells (and the upward steps found near the ends of the energy trough for the $3_{10}$-helical model, Fig.~\ref{fig:Ebothhelices}F) are a collective result of the entire charge cluster in the VS interacting with induced charge on the protein boundary. The more dense cluster of the $3_{10}$-helical model has stronger interactions among charges and counter-charges, and with the charges induced on the protein dielectric, than does the more spread-out (in the angular dimension) cluster of charges in the α-helical model. The difference is strongest in the extreme allowed positions of S4 travel, when a maximal number of S4 charges is located in either one of the aqueous phases near the protein boundary.

These simulations with varied dielectric coefficient thus lead us to
conclude that protein polarizability is not crucial for electrostatic
stability of a sliding-helix VS that balances S4 charges by
counter-charges. On the other hand, the degree of polarizability
contributes to determining the relation between voltage and displaced
charge.

\subsection{Geometry of the gating canal}

About 10 residues of the S4 helix are inaccessible to hydrophilic
reagents applied to resting channels from either the extracellular or
intracellular side \citep{gandhi:2002}. If the S4 domain is
α-helical and oriented perpendicular to the membrane plane,
then the inaccessible translational range is only ≈1.35~nm
long and thus shorter than the lipid core of the membrane. Therefore
the S4 helix is thought to move across the hydrophobic region in a
short `gating canal' that forms an electrical seal between water
pockets extending toward the canal from either side (vestibules). One
consequence of this arrangement is that most of the applied electrical
voltage drops over a short length of the gating canal and is thereby
`focused' on the interval of S4 charges dwelling in the canal itself.

Short of a simulation, approximate inferences can be made of potential
electrostatic consequences of the gating canal. If the queue of S4
charges spans the gating canal plus the range of S4 travel, then the instantaneous 
force $F$ exerted by the applied field on the $n$ S4 charges
simultaneously present in the canal of length $D$ is approximately $F=
n\text{e}_0V_m/D$. The length $D$ and the number $n$ are related by the
axial spacing of the charges in the S4 helix: $d=D/n$ (assuming that
`in the canal' on average implies an extra distance at either end of
about $d/2$). The electrical force acting on the S4 charges is thus
$F=\text{e}_0V_m/d$. Therefore this force $F$ is independent of the length
of the gating canal, depending only on the spacing between S4
charges. The instantaneous work done by the field on the S4 charges as the S4 moves
through the canal is given by $W=Fmd=m\text{e}_0V_m$, where $m$ is the
number of S4 charges that travel across one of the gating canal mouths
during the motion associated with the work $W.$ This work does not depend
on the length of the gating canal. Moreover, the S4 charges travel the
same distance $md$ per unit work regardless of the length of the
gating canal, $D$. Voltage sensitivity (defined by force or work
picked up per unit applied voltage) does not depend on the focusing of
the electric field by the gating canal itself, as long as the queue of
S4 charges extends across the gating canal.

\begin{figure}
  \centering
  \includegraphics{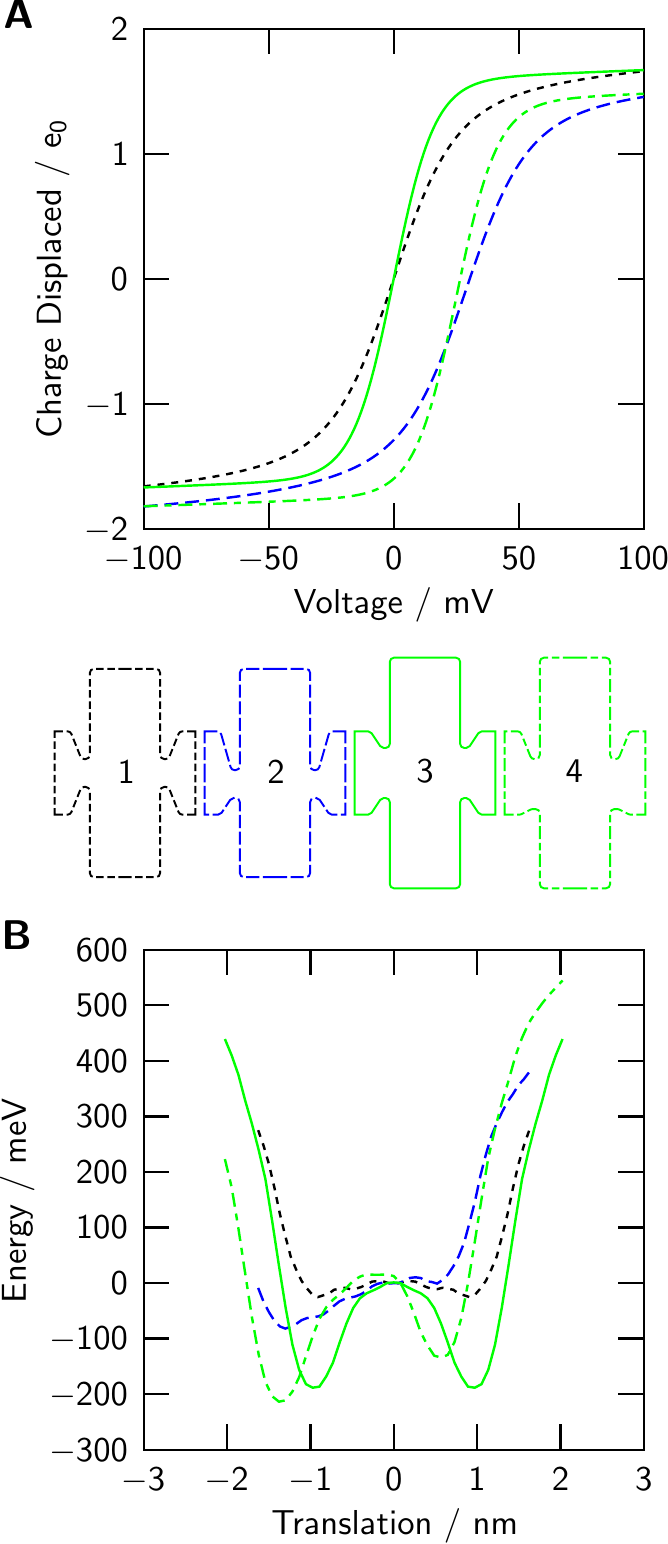}
  \caption{\emph{Consequences of gating pore geometry.} Length and/or
    position of the gating canal (and hence its vestibules) are varied
    as shown by the outlines of the VS protein between the
    graphs. Outline~1 is our standard α-helical S4 model. The
    line patterns \& colors in both the graphs of the charge/voltage
    relations (A) and the underlying translational energy profiles (B)
    correspond to the line patterns \& colors used for the outlines of
    the VS shapes. Table~\ref{tab:cell} gives further geometrical
    specifications for shapes 1--4. Outline~4 can be seen in the
    3-dimensional Fig.~\ref{fig:movie:asymm}.}
  \label{fig:gatpore}
\end{figure}

To assess more fully the consequences of gating canal geometry,
we simulate varied canal lengths and/or canal positions with respect
to the membrane planes in the VS model
(Fig.~\ref{fig:gatpore}). Starting with the geometry of our standard
model for the α-helical VS (VS protein outline 1, compare
Fig.~\ref{fig:cell}), we shift the canal toward the intracellular
membrane plane (outline 2). Two further variations include a reduction
of the vestibule length on both sides by one half, thereby lengthening
the gating canal (outline 3) and a shift of that lengthened gating
canal (outline 4, Fig.~\ref{fig:movie:asymm}).

With both the centered and the shifted gating canals, lengthening the
canal increases the slope of the charge/voltage relation while
maintaining the maximal charge displacement between $\pm$100~mV
(Fig.~\ref{fig:gatpore}A, line patterns in the graph correspond with
the protein outlines below the graph). Shifting the position of the
gating canal shifts the charge/voltage relation along both the voltage
axis and the charge axis (towards the intracellular side) while
maintaining the slope typical of the canal length. These shifts
indicate that asymmetries of the vestibules of the gating canal bias
the voltage sensed by the S4 segment.

The underlying energetics are summarized in
Fig.~\ref{fig:gatpore}B. Lengthening the gating canal increases the
depth of the energy minima at the ends of the range of travel, thus
promoting the bistable behavior of the VS associated with a steep
slope of the charge/voltage relation. Shifting the position of the
gating pore shifts the energy profile and therefore the S4 geometrical
range of travel. The translational spacing between the minima at
either end of the energy trough is not changed by these variations of
gating canal geometry.

The observed invariance of the total charge displaced in these
simulations conforms to our earlier approximation
($W=m\text{e}_0V_m$). Geometrical S4 travel relative to the center of
the membrane tends to follow the position of the gating canal. A
longer gating canal giving \emph{less} `focusing' of the applied
electric field results in a \emph{steeper} charge/voltage
relation. The latter two relations are due to the self-energy
contribution to the configurational energy as defined by
Eq.~\ref{eq:nullenergy}, that is to the electric field when
$V_{\mathrm{m}}=0$ is applied to the bath electrodes. In this
section's simulations, the counter-charges to the S4 charges are
aligned to the center of the septum of the gating canal which, when
the gating canal is shifted, varies from the center of the membrane
and the S4 protein region. The counter-charges' mutual spacing
continues to be constant. With a longer canal, there is a longer gap
between the outermost counter-charge and the central end of the
vestibule on each side. The energy minima of the translational energy
profile are located at the S4 translational coordinate associated with
the trailing charge of the S4 helix vacating the poorly screened
stretch of the gating canal (see Animation~\ref{movie:1-2-3-8-9-0} for
associated charge distributions).

\subsection{Screening of VS charge exposed to the aqueous solutions}

Translocation of S4 charge across the membrane necessarily shifts a
high local density of protein charge across the membrane. Therefore
the energetics of S4 translation should depend on the screening on
either side of the membrane of S4 charges by bath ions. We have thus
far reported on VS models under symmetrical screening conditions;
under those conditions, any screening effects on S4 energetics are
masked by mutual compensation. Strongly asymmetrical screening should
unmask the contribution of screening to the energetics. Our simulation
does not include explicit bath ions, as we explore screening effects
instead by varying the distance of the bath electrodes from the
membrane and VS protein surfaces (see section `Model and boundary
conditions'). The two extreme electrode placements shown in
Fig.~\ref{fig:cell}A,C are equivalent to low millimolar ionic strength
baths and exceedingly large ionic strength baths.

\begin{figure}
  \centering
  \includegraphics{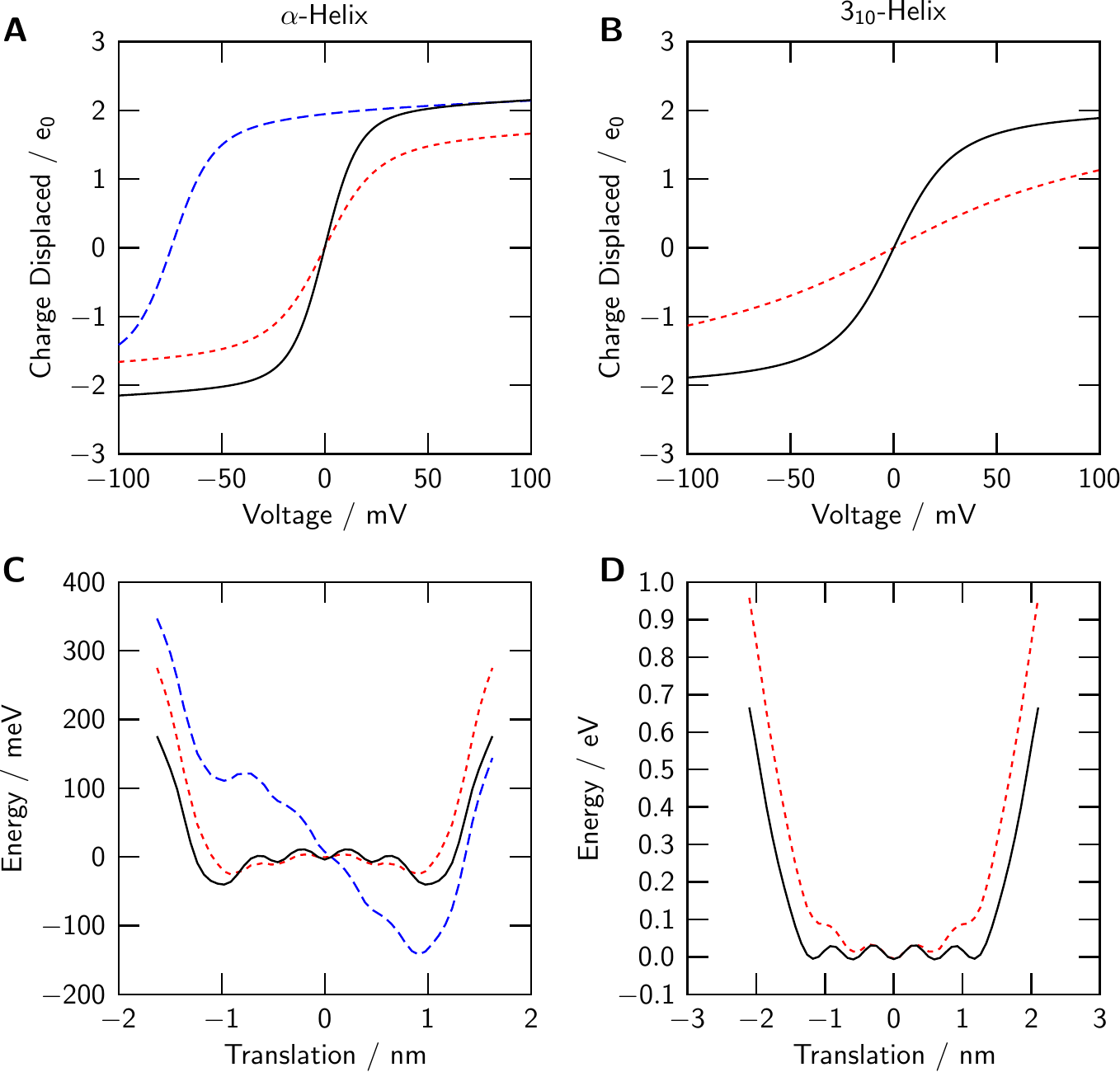}
  \caption{\emph{Consequences of screening by the baths.} (A, C) The
    standard α-helical S4 model is simulated with varied
    electrode arrangements to mimic varied ionic strength of the baths
    (see Fig.~\ref{fig:cell}). Charge/voltage relations and
    translational energy profiles correspond to symmetrical low ionic
    strengths (short-dashed lines, geometry as in
    Fig.~\ref{fig:cell}A), symmetrical high ionic strengths (solid
    lines, geometry as in Fig.~\ref{fig:cell}C), and external low and
    internal high ionic strengths (long-dashed line, geometry as in
    Fig.~\ref{fig:cell}B). (B, D) Analogous simulations of the
    $3_{10}$-helical S4 model for symmetrical low or high ionic
    strengths.}
  \label{fig:bathorel}
\end{figure}

We compute the charge/voltage relations for our standard
α-helical model with three electrode configurations
(Fig.~\ref{fig:bathorel}A): (1) both electrodes far from the membrane
and protein, (2) both electrodes directly on the surface of the
membrane and protein, and (3) the external electrode on the membrane
and protein, and the internal electrode far from the membrane and
protein. The charge/voltage relation for configuration~2 (solid line)
reveals a larger total displaced charge and a steeper slope compared
with configuration~1 (dotted line). This increase in total charge at
`high ionic strength' has two sources: the increase in electric
distance between geometrical S4 positions (as discussed by
\citealp{islas:2001}); and the increase in the S4 geometrical range of
motion at $\pm$100~mV, as indicated by the energy minima of
translational energy profiles (Fig.~\ref{fig:bathorel}C). The increase
in slope is due to the deeper translational energy minima at the
extreme positions, which promote bistable behavior as described
earlier.

The simulation for configuration~3 (low internal and high external
ionic strength) yields a left-shifted charge/voltage relation
(long-dashed line in Fig.~\ref{fig:bathorel}A). The magnitude of this
shift is due to the asymmetrical screening of S4 charges on the
internal and external sides; the S4 charges are attracted toward the
external bath where screening is more effective. This result shows
that the screening of bath-exposed S4 charges contributes
substantially to the energetics of S4 translation. This contribution
is masked by mutual compensation when ionic strengths are equal on
both sides, except for contributions in terms of more subtle variation
of the shape of the charge/voltage relation.

The $3_{10}$-helical model gives substantially smaller voltage
sensitivity than the α-helical model when we simulate the
models with `low ionic strength' electrode placement. Given the
effects of screening on voltage sensitivity seen in
Fig.~\ref{fig:bathorel}A, it is interesting to see to what extent
strong screening `rescues' voltage sensitivity of the $3_{10}$-helical
model. In fact, exceedingly large ionic strength increases voltage
sensitivity substantially (solid line in Fig.~\ref{fig:bathorel}B) in
terms of both increased maximum slope and total charge displaced
across the simulated potential. However, this `rescue' does not push
voltage sensitivity all the way to that found for the α-helical
model (solid line in Fig.~\ref{fig:bathorel}A). The corresponding
translational energy profiles for the $3_{10}$-helical model are shown
in Fig.~\ref{fig:bathorel}D.

\citet{islas:2001} found small voltage shifts (<10~mV) when studying
the effects of intracellular variation of ionic strength on
charge/voltage relations of \emph{Shaker} channels. They interpreted
these observations in terms of the Guy-Chapman theory of the
electrical double layer. In those terms, those small shifts indicate a
low density of membrane surface charge. In our simulations
(Fig.~\ref{fig:bathorel}A), a unilateral variation of screening causes
a large shift even though the model does not include surface
charge. Screening of S4 charges by a bath of low ionic strength
reduces the electrostatic energy less than the screening of such S4
charges by a bath of high ionic strength. This effect does not require
fixed surface charges, but arises from the S4 charges themselves.  In
contrast to the interpretation given by \citet{islas:2001}, the
absence of a voltage shift may indicate that fixed negative charge
exists at the intracellular end of the gating canal, but that such
charge is largely neutralized by the positive S4 charges exposed to
the intracellular baths at the midpoint of the S4's range of
travel. The S4 charge exposed in this configuration is then screened
by the negative fixed charge regardless of the ionic strength of the
bath.

\subsection{Consequences of `surface charge'}

The existence of `surface charge' acting on the VS was inferred by
\citet{frankenhaeuser:1957} and many subsequent studies
\citep{hille:2001}. Negative charges on residues of the extracellular
loop connecting the S5 and pore segments of Kv channels determine the
voltage for activating the channel
\citep{elinder:1998,elinder:1999}. One residue is conserved in that
region -- E418 for \emph{ShakerB} -- while others are variable. E418
is involved in mediating activation-related slow inactivation and has
been proposed to interact at close range with the outermost S4 charge
(R362) when the VS is in the activating position
\citep{larsson:2000,elinder:2001}.

We model surface charge as discrete by placing a point charge of
-$\text{e}_0$ into either one bath at translation 1~nm or both
baths at $\pm$1~nm, which are the translations approached by the
terminal S4 charges at voltages of $\pm$100~mV. The charge is given
the same radial distance from the S4 axis as the counter-charges
located inside the protein. Two angular positions are tested: a
`close' one aligned with the terminal S4 charge, and a `far' one
turned 180\textdegree away from the S4 charge.

\begin{figure}
  \centering
  \includegraphics{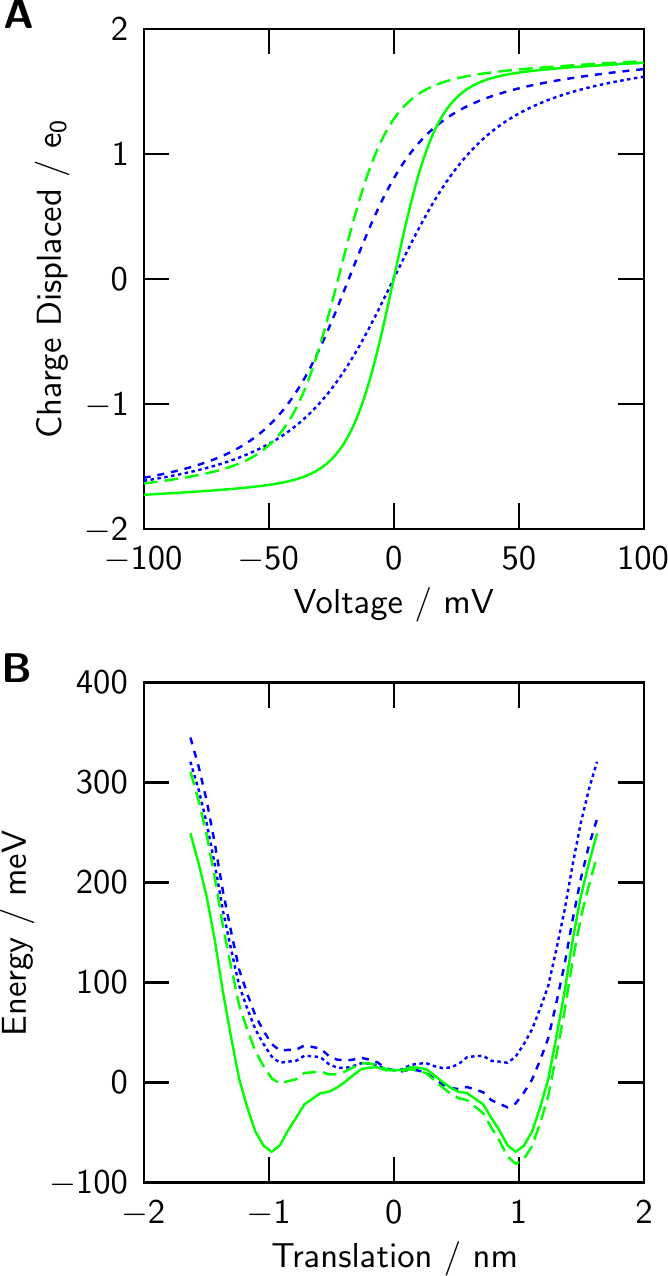}
  \caption{\emph{Consequences of `surface charge'.} Standard
    α-helical S4 model with discrete surface charge(s) added in
    three variations of geometrical position (detailed in text): close
    external \emph{and} internal surface charges (\emph{solid line});
    close external-only surface charge (\emph{long-dashed line}); far
    external \emph{and} internal surface charges (\emph{dotted line}; far
    external-only surface charge (\emph{short-dashed line}). (A)
    Statistical-mechanical expectations of charge/voltage relations;
    (B) Translational energy profiles.}
  \label{fig:alphasigma}
\end{figure}

The charge/voltage relations for our standard α-helical S4 model with
four surface charge arrangements added are shown in
Fig.~\ref{fig:alphasigma}A: one external charge in the far position
(\emph{short-dash line}); one external and one internal surface
charge, with both in the far positions (\emph{dotted line}); one
external charge in the close position (\emph{long-dash line}); and one
external and one internal charge, with both in the close positions
(\emph{solid line}). The far external charge shifts the charge/voltage
relation, producing the signature of a surface charge. The combination
of far external and far internal surface charges slightly reduces the
slope of the charge/voltage relation with respect to the model without
surface charges (compare Fig.~\ref{fig:qvsbothhelices}). The
combination of close internal and close external charge also produces
no shift (due to symmetry) but does steepen the slope of the
charge/voltage curve (compare Fig.~\ref{fig:qvsbothhelices}A). With
only the close external charge, the charge/voltage relation is
shifted, and at negative voltage, the asymptote is slightly less steep
relative to the asymptote with both surface charges. Thus surface
charges have position-dependent effects on both the midpoint and slope
of charge/voltage relations.

\citet{elinder:2001} estimated the contributions of
individual surface charges present in $\text{K}^+$ channels to voltage shift
in charge/voltage relations, using the voltage shifts of ionic
conductance observed with varied $\text{Mg}^{2+}$ concentrations. Our
simulated shifts are in the range of those estimated
shifts. \citet{elinder:2001} interpreted the varied shifts estimated
for different charged residues to be a consequence of the distance
from the VS charges to the charged residues. Our self-consistent
simulations with a `close' and a `far' surface charge show only a
small effect on voltage shifts due to the distance to the surface
charge (long- and short-dashed lines in Fig.~\ref{fig:alphasigma}).

\ifthenelse{\boolean{internal}}{\enlargethispage{\baselineskip}}{} The
translational energy profiles from those four surface-charge
simulations are shown in Fig.~\ref{fig:alphasigma}B. Far external
surface charge chiefly tilts the profile (\emph{short-dash line}),
whereas far surface charges on both sides remove the energy wells near
the ends of the energy trough. Close external and internal charges,
however, create deeper energy wells at either end of the range of S4
motion (\emph{solid line}). With only a close external charge, an
energy well is added at the external end of the range of S4 motion,
while energy at the intracellular end of the range is raised
(\emph{long-dash line}). The energy well resulting from close
interaction with a surface charge increases the electrostatic
contribution to S4 transmembrane stability as well, since it lowers
the electrostatic energy of that S4 position relative to more extreme 
S4 positions. This does not occur when the interaction with the
surface charge is far.

\section{Conclusion}

The sliding helix configuration has an innate electrostatic stability
if the helix charges not exposed to water are balanced by fixed
counter-charges. This balance extends over the range of helix movement
where from about 3 to about 4 charge equivalents are displaced
across the membrane field. The limited range of motion at the extremes
of physiological membrane potential leads to robustness of VS behavior
over a range of conditions and geometrical variations. Whether the
geometry is α-helical, $3_{10}$ helical, in high ionic
concentration or low, this basic structural issue leads to a system
that results in gating-charge displacement in response to change in
transmembrane potential.

On the other hand between those extremes, the energy landscape for a
given membrane potential varies between models: from almost flat to
two energy troughs separated by a large barrier. Those variations lead
to distinct charge displacement patterns (implying distinct kinetics
not investigated here). Protein dielectric, gating canal geometry,
helix configuration and surface charges alter the behavior of the
system, ranging from systems that can be approximated by two-state
Boltzmann models all the way to continuous model systems. This results
in a family of voltage-sensors with easily perturbed voltage
sensitivity despite their robust behavior at their limits. Features
not made explicit in our models will likely play a role in that region
of sensitivity.

These results suggest a problem with the $3_{10}$-helical S4
structure: in the parameter space we studied, the $3_{10}$ models tend
towards poor voltage-sensitivity in contrast with α-helical
models which show both higher voltage-sensitivity and greater
variation in voltage-sensitivity.

Our simulations exploring the consequences of a set of physical
parameters elucidate the electrostatic principles underlying both the
robustness of voltage-sensor charge-displacement response and the
sensitivity of voltage-sensitivity and gating charge voltage shift in
the face of small model variations. Resolving the `inverse' problem of
connecting specific experimental data with the specific underlying
physics is both a computational and experimental challenge. We suggest
that more `forward' simulations of the consequences of physical
parameters are needed to guide the design and interpretation of
experiments intended to determine fine control of voltage sensing.

\section{Acknowledgments}

The authors are grateful for the support of the National Institutes of Health (grant GM083161) to W. Nonner and a Graduate Research Fellowship of the National Science Foundation to A. Peyser. We thank Drs. Alice Holohean, Peter Larsson, and Karl Magleby  for helpful discussions.

\newpage
\begin{flushleft}
\bibliography{mode=abbrv}
\end{flushleft}

\supplementary
\subsection{Tests of numerical accuracy in solving the electrostatics.}

The divergence theorem (Gauss's theorem) states that
\begin{equation}
\oint_{\mathcal{S}}\!\epsilon\left(\mathbf{r}\right)\epsilon_0
   \mathbf{E}\left(\mathbf{r}\right) \cdot 
   \mathbf{n}\left(\mathbf{r}\right)\;da
   = \int_{\mathcal{V}}\!\rho^{\text{src}}\left(\mathbf{r}\right)\;d\tau
   \label{eq:gauss}
\end{equation}
where $\mathcal{S}$ is the closed surface around the volume
$\mathcal{V}$, $da$ is the area of the surface element located at
$\mathbf{r} \in \mathcal{S}$, and $d\tau$ is the volume of the space
element located at $\mathbf{r} \in \mathcal{V}$. This holds for any
geometry of $\mathcal{V}$.

A relevant surface for applying Gauss's theorem is the dielectric
boundary of the VS protein since the electric field at that boundary
generates the most important induced charge to be computed. The
surface integral in Eq.~\ref{eq:gauss} must yield the algebraic sum of
the formal charges that are assigned to the VS residues. Our numerical
method approximates the normal field strength (and hence induced
charge density) at the surface as piecewise uniform. Gauss's theorem
provides a test in the actual system that we simulate of the accuracy
of the induced charge computed on the boundary. The difference between
the computed total induced charge $Q_{\text{calc}}$ and the charge
$Q_{\text{Gauss}}$ predicted from Gauss's theorem is:
\begin{equation}
  Q_{\text{calc}} - Q_{\text{Gauss}}  = 
  -\sum_j \frac{\epsilon_{\mathrm{p}} \epsilon_j}{\epsilon_j - \epsilon_{\mathrm{p}}}
          \sigma^{\text{ind}} a_j - \sum_k \dbl{q}{k}{\text{src}}
\label{eq:gausserror}
\end{equation}
where $a_j$ is the area of protein surface element $j$ and
$\epsilon_j$ is the dielectric coefficient on the out-facing side of
$a_j$.

The error in the computed charge calculation is likely to vary as S4
charges are moved in a simulation since the distances between the S4
charges and the dielectric boundaries vary. Fig.~\ref{fig:gausstest}
shows the error in induced charge calculation for the full range of S4
translational positions in a typical simulation. The error is
≤0.008~$\text{e}_0$ of the actual net charge of 3~$\text{e}_0$ assigned to the VS
in this simulation. The results of our charge calculation are thus in
good agreement with Gauss's theorem, as they must be.

\begin{figure} \centering
  \includegraphics{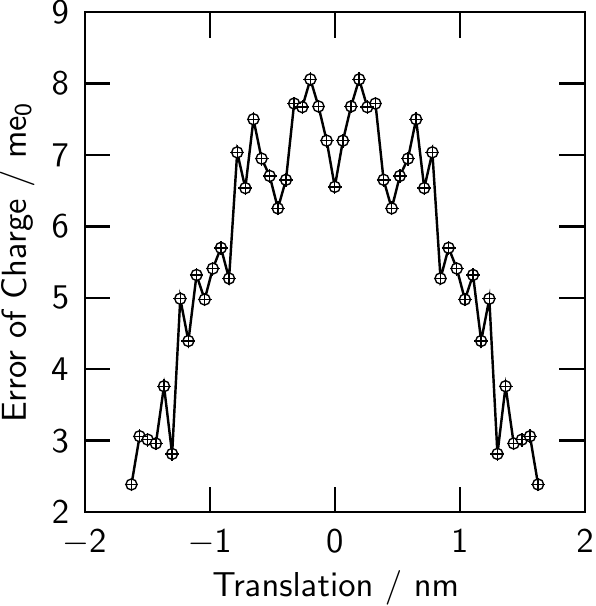}
  \caption{\emph{Test of numerical accuracy using Gauss's theorem.}
    The difference $Q_{\text{calc}}-Q_{\text{Gauss}}$ as defined
    by Eq.~\ref{eq:gausserror} is plotted versus S4 translation
    (Eq.~\ref{eq:randvarrot} is applied to calculate the expectation
    value over the rotational degree of freedom).  Normal electric
    flux is integrated over the surface of the protein (brown region
    in Fig.~\ref{fig:cell}). The protein region contains a net charge
    of +3~$\text{e}_0$ (6 positive S4 charges and 3 negative counter-charges),
    which is recovered from the integrated electrical flux within a
    relative error of <0.003 over the full range of S4 translation
    (≤0.008~$\text{e}_0$). Charge errors computed for three applied voltages
    (-100~mV, circles; 0~mV, line; 100~mV, crosses) are superimposed
    in the graph, showing no detectable effect of the external field
    on the error.}
  \label{fig:gausstest}
\end{figure}

In a second control, we compare the calculations of electrostatic
energy by two independent approaches: (1) direct calculation of the
energy, using the Ramo-Shockley theorem as described in Methods as our
standard procedure; and (2) path-integration of the force acting on
the S4 segment via Eq.~\ref{eq:ework2}. Both methods use the induced
charges on dielectric boundaries and electrodes computed in our
simulation and tested using Gauss's law as described above. Method (1)
is based on the potentials produced by all charges. Method (2) is
based on the electrical fields of all charges. These methods have no
computational overlap following the charge computation (in other
words, these methods use distinct code paths, a test for coding
correctness). We show this control for three applied voltages with a
prescribed (diagonal) path through the translational and rotational
dimensions of the range of S4 motion which we typically
examine. Fig.~\ref{fig:Ebypath} shows the energies from Method (1) as
dots superimposed on the energies from method (2) as lines.

\begin{figure}
  \centering
  \includegraphics{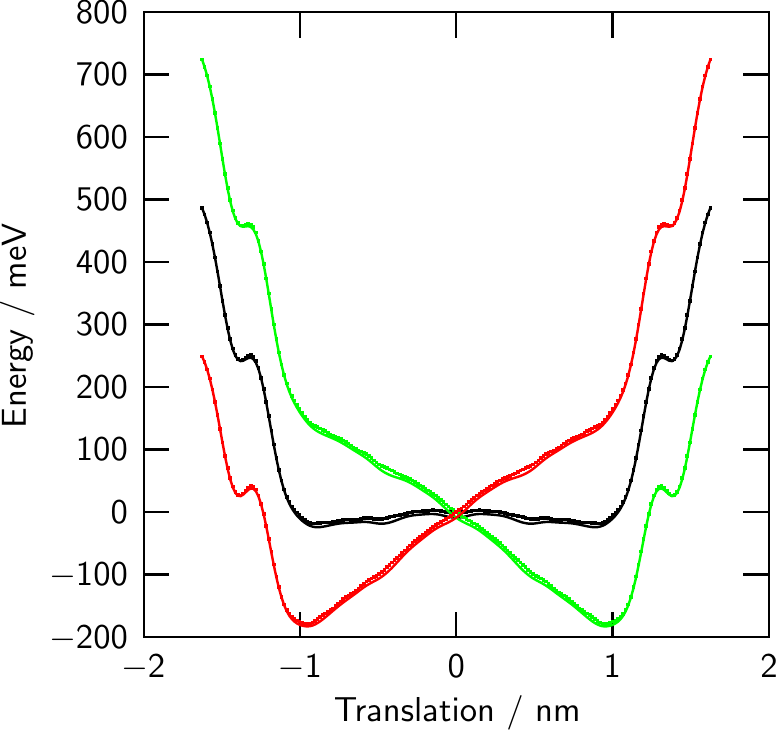}
  \caption{\emph{Control for the computation of energy.} Energy
    profiles for the fixed path prescribed by the translations plotted
    as abscissa and the proportional rotational increments such that
    the standard α-helical model S4 charges follow the curve of
    counter-charges (from -180 to +180 degrees). Energies computed by
    integrating the force along this path (lines) are compared with
    energies computed using the Ramo-Shockley theorem (dots) as
    described as our routine method in Methods. These tests are
    applied for three membrane voltages (100~mV, red; 0~mV, black;
    +100~mV, green). The energies obtained by these two methods are
    aligned at the most negative translation tested.}
\label{fig:Ebypath}
\end{figure}

\subsection{Charge distributions of VS models}

The sliding helix in our VS models is a microscopic voltage sensor,
thus subject to thermal agitation. Its behavior must be described in
statistical terms. In this section, stochastic VS behavior is
visualized for several models presented in the paper, both as figures
for a fixed applied voltage (-100~mV;
Figs.~\ref{fig:movie:first}--\ref{fig:movie:last}) and in animations
with voltage increasing uniformly over time from -100 to +100~mV
(Animations~\ref{movie:first}--\ref{movie:last}, respectively).

The figures show two stochastic aspects of VS behavior: (1) the mean
positions of the S4 charges (marked by blue balls), and (2) the charge
density distribution of S4 charge (represented by a blue cloud with a
color intensity proportional to the charge density there). A high
density of color marks the locations where the S4 charges dwell
frequently, as opposed to their mean positions.

The mean position for each charge is computed from
Eq.~\ref{eq:randvar} using the positions $\mathbf{r}_k$ for the charge
$q_k$ as the random variable $X$, using the partition function with
translational and rotational degrees of freedom.  Since the helix
behaves as a solid body, the helix position $\mathbf{r}$ fixes the
positions $\mathbf{r}_k$ for the charges $q_k$. That relationship
allows us to define the partition function for the positions
$\mathbf{r}_k$ in terms of the partition function for the helix
position $\mathbf{r}$, as well as the energy functions for
$\mathbf{r}_k$ in terms of the energy function for the helix position.

In other words, by applying Eq.~\ref{eq:randvar} to a model with $X =
\mathbf{r}_k$, the mean position $\langle\mathbf{r}_k\rangle$ for any
charge can be calculated given the probability of configuration $ij$
($P_{ij}$, Eqs.~\ref{eq:dist} \& \ref{eq:partitionfct}):
\begin{equation}
  \langle\mathbf{r}_k\rangle = \sum_{i,j} \mathbf{r}_{ijk}P_{ij}
  = \frac{1}{\mathcal{Q}} \sum_{i,j} \mathbf{r}_{ijk} e^{-W_{ij}/k_BT}
  \label{eq:pos:mean}
\end{equation}
where $W_{ij}$ is the work to construct configuration $ij$ and
$\mathbf{r}_{ijk}$ is the position of charge $k$ in configuration
$ij$.

Likewise, the distribution of charge can be computed by applying the
partition and energy functions in terms of the position of charges
$\mathbf{r}_k$. The charge density $\bar z(\mathbf{r})$ is then
the sum over all charges of the probability of each charge being
located at $\mathbf{r}$, multiplied by its valency, and normalized:
\begin{equation}
  \bar z(\mathbf{r}) 
  = \frac{1}{\sum\limits_{i,j,k} P_{ij}z_k}
    \sum_{i,j,k} P_{ij}z_k\,\delta(\mathbf{r},\mathbf{r}_{ijk})
  \label{eq:pos:prob}
\end{equation}
where $\delta(\mathbf{r}, \mathbf{r}_{ijk})$ is the discretized delta
function (1 if we're treating $\mathbf{r}_{ijk}$ as the same
location as $\mathbf{r}$ for visualization purposes, otherwise
0). The color representations for the animations are proportional to
$\bar z(\mathbf{r})$ normalized to the highest charge density at that
frame's potential.

\begin{figure}
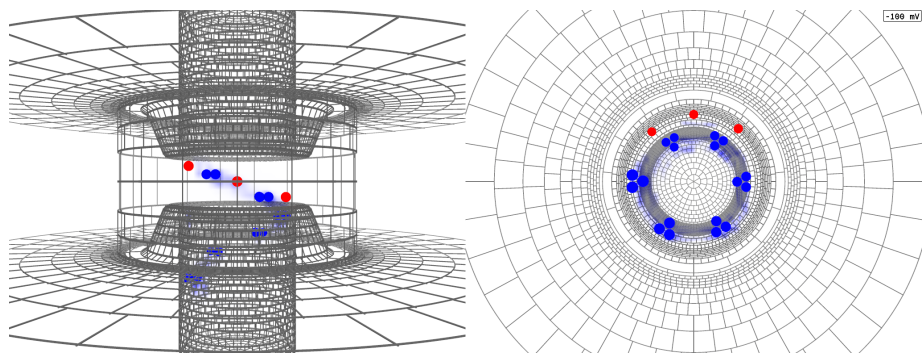

  \centering
  \moviegraphic{1-2-3-8-0-0}
  \caption{\emph{Standard \emph{α}-helical model:
      \attachmovie[\label{movie:first}]{1-2-3-8-0-0}}. Position and
    distribution of charges for model α~(1) in
    Table~\ref{tab:cell:params} with $\epsilon_p=4$. Red symbols
    represent fixed negative counter-charges, blue symbols represent
    the mean position of -1/3~e$_0$ on S4 arginines, and blue
    shading represent the relative probability of negative charge at a
    given position. This is the model used in Fig.~\ref{fig:qvsbothhelices}A.}
  \label{fig:movie:standard}
  \label{fig:movie:first}
\end{figure}

\begin{figure}
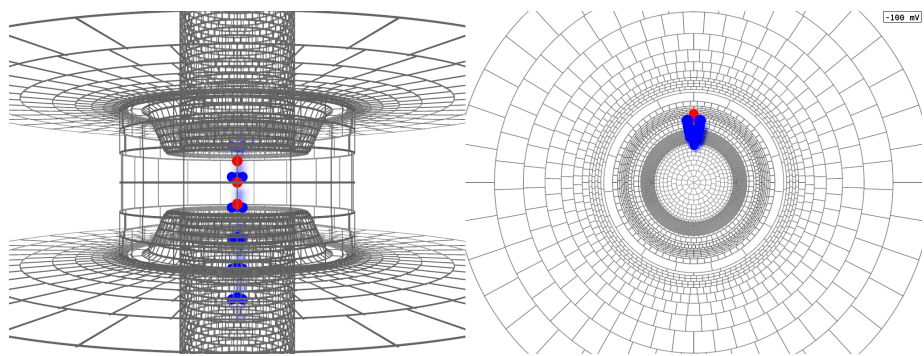

  \centering
  \moviegraphic{1-2-20-41-0-0}
  \caption{\emph{Standard $3_{10}$-helical model: \attachmovie{1-2-20-41-0-0}}.
    Position and distribution of charges for model $3_{10}$ in
    table~\ref{tab:cell:params} with $\epsilon_p=4$. See the
    description of Fig.~\ref{fig:movie:first} for further details.
    This is the model used in Fig.~\ref{fig:qvsbothhelices}B.}
  \label{fig:movie:310}
\end{figure}

\begin{figure}
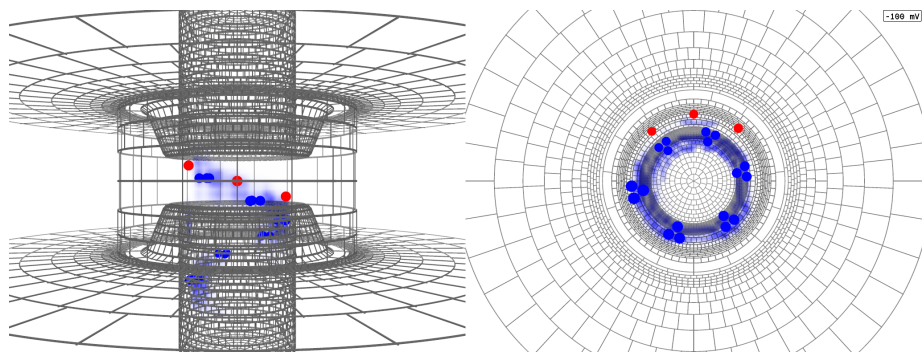

  \centering
  \moviegraphic{1-7-3-8-0-0}
  \caption{\emph{\emph{α}-helical model with $\epsilon_p=16$:
      \attachmovie{1-7-3-8-0-0}}. Position and distribution of
    charges for model α~(1) in table~\ref{tab:cell:params} with
    $\epsilon_p=16$. See the description of Fig.~\ref{fig:movie:first}
    for further details. This is the model used in Fig.~\ref{fig:epsilon}.}
  \label{fig:movie:eps16}
\end{figure}

\begin{figure}
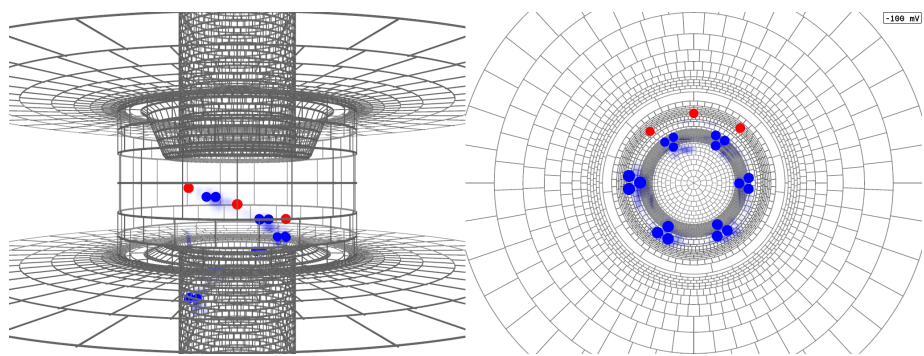

  \centering
  \moviegraphic{1-2-3-8-9-0}
  \caption{\emph{\emph{α}-helical model with asymmetric gating canal:
      \attachmovie[\label{movie:last}]{1-2-3-8-9-0}}.  Position and
    distribution of charges for model α~(4) in
    table~\ref{tab:cell:params} with $\epsilon_p=4$. See the
    description of Fig.~\ref{fig:movie:first} for further details.
    This is the model used in Fig.~\ref{fig:gatpore}, outline 4.}
  \label{fig:movie:asymm}
  \label{fig:movie:last}
\end{figure}

\end{document}